\documentclass[a4paper,11pt]{article}
\pdfoutput=1
\usepackage{jheppub}
\usepackage[T1]{fontenc}
\usepackage{amsmath}
\usepackage{hyperref}
\usepackage{color}
\usepackage{graphicx}
\usepackage{amsfonts,amscd,dsfont}
\usepackage{bbold}
\usepackage{amsthm}
\usepackage{amssymb}
\usepackage{subfigure}

\newcommand{\eps}{\epsilon}
\newcommand{\ord}{\begin{cal}O\end{cal}}

\def\be{\begin{equation}}
\def\ee{\end{equation}}
\def\beq{\begin{equation}}
\def\eeq{\end{equation}}
\def\bsp#1\esp{\begin{split}#1\end{split}}

\newcommand{\cA}{\begin{cal}A\end{cal}}

\newcommand{\cL}{\begin{cal}L\end{cal}}

\newcommand{\cO}{\begin{cal}O\end{cal}}

\newcommand{\cR}{\begin{cal}R\end{cal}}

\newcommand{\cl}{\textrm{Cl}}

\title{Gluon-fusion Higgs production in the Standard Model Effective Field Theory}

\author[a,b]{Nicolas Deutschmann}
\author[b,c]{, Claude Duhr}
\author[b]{, Fabio Maltoni}
\author[d]{, Eleni Vryonidou}
\affiliation[a]{Univ. Lyon, Universit\'e Lyon 1, CNRS/IN2P3, IPNL, F-69622, Villeurbanne, France}
\affiliation[b]{Centre for Cosmology, Particle Physics and Phenomenology (CP3), \\ Universit\'e catholique de Louvain, Chemin du Cyclotron 2, 1348 Louvain-La-Neuve, Belgium}
\affiliation[c]{Theoretical Physics Department, CERN, CH-1211 Geneva 23, Switzerland}
\affiliation[d]{Nikhef, Science Park 105, 1098 XG, Amsterdam, The Netherlands}
\emailAdd{n.deutschmann@ipnl.in2p3.fr}
\emailAdd{claude.duhr@cern.ch}
\emailAdd{fabio.maltoni@uclouvain.be}
\emailAdd{eleni.vryonidou@nikhef.nl}

\preprint{\begin{flushright}CP3-17-24 \\CERN-TH-2017-165 \\NIKHEF-2017-035\end{flushright}}

\abstract{We provide the complete set of predictions needed to achieve NLO accuracy in the Standard Model Effective Field Theory at dimension six for Higgs production in gluon fusion. In particular, we compute for the first time the contribution of the chromomagnetic operator $ \bar Q_L \Phi  \sigma q_R G$ at NLO in QCD, which entails two-loop virtual and one-loop real contributions, as well as renormalisation and mixing with the Yukawa operator $\Phi^\dagger \Phi\,  \bar Q_L  \Phi  q_R$ and the gluon-fusion operator $\Phi^\dagger \Phi\, GG$.  Focusing on the top-quark-Higgs couplings, we consider the phenomenological impact of the NLO corrections in constraining the three relevant operators by implementing the results into the {\sc MadGraph5\_aMC@NLO}  framework. This allows us to compute total cross sections as well as to perform event generation at NLO  that can be directly employed in experimental analyses. }

\begin{document}
\maketitle


\section{Introduction}
\label{sec:intro}

Five years into its discovery at the LHC, the Higgs boson is still the centre of attention of the high-energy physics community.  A wealth of information has been collected on its properties by the ATLAS and CMS experiments~\cite{Khachatryan:2014qaa,Aad:2014lma,Aad:2015iha,Aad:2015gra,Khachatryan:2016vau}, all of which so far support the predictions of the Standard Model (SM). In particular, the size of the couplings to the weak vector bosons and to the electrically charged third generation fermions has been confirmed, and the first evidence of the coupling to second generation fermions (either charm quark or muon) could arrive in the coming years, if SM-like. 

The steady improvement in the  precision of the current and forthcoming Higgs measurements invites to explore physics beyond the SM not only via the search of new resonances, as widely pursued at the LHC, but also via indirect effects on the couplings of the Higgs boson to the known SM particles. The most appealing aspect of such an approach is that, despite being much more challenging than direct searches both experimentally and theoretically,  it has the potential to probe new physics scales that are beyond the kinematical reach of the LHC.  A powerful and predictive framework to analyse  possible deviations in the absence of resonant BSM production is provided by the SM Effective Field Theory (SMEFT) \cite{Weinberg:1978kz,Buchmuller:1985jz,Leung:1984ni}, {i.e.},~the SM augmented by higher-dimensional operators. Among the most interesting features of this framework is the possibility to compute radiative corrections in the gauge couplings, thus allowing for systematic improvements of the predictions and a reduction of the theoretical uncertainties~\cite{Passarino:2012cb}. In particular, higher-order corrections in the strong coupling constant typically entail large effects at the LHC both in the accuracy and the precision. They are therefore being calculated for a continuously growing set of processes involving operators of dimension six featuring the Higgs boson, the bottom and  top quarks and the vector bosons. Currently, predictions for the most important associated production channels for the Higgs boson are available in this framework, {e.g.}, VH, VBF and $t\bar tH$~\cite{Maltoni:2013sma, Degrande:2016dqg, Maltoni:2016yxb}. For top-quark production, NLO results for EW and QCD inclusive production, {i.e.}, $tj$ and $t\bar t$, and for top-quark associated production $t\bar tZ$, $t\bar t\gamma$ have also appeared \cite{Degrande:2014tta,Franzosi:2015osa,Zhang:2016omx,Bylund:2016phk,Rontsch:2014cca,Rontsch:2015una}. The effect of dimension-six operators has also become available recently for top-quark and Higgs decays~\cite{Zhang:2014rja, Hartmann:2015oia,Ghezzi:2015vva,Hartmann:2015aia,Gauld:2015lmb}. 

The situation is somewhat less satisfactory for gluon fusion, which, despite being a loop-induced process in the SM, is highly enhanced by the gluon density in the proton and provides the most important Higgs-production channel at the LHC.   In the SM, the QCD corrections are now  known up to N$^3$LO in the limit of a heavy top quark~\cite{Anastasiou:2014lda, Anastasiou:2015ema,
Anastasiou:2016cez}. The full quark-mass dependence is known
up to NLO \cite{Graudenz:1992pv, Spira:1995rr, Harlander:2005rq,
Anastasiou:2009kn}, while at NNLO only subleading terms in the heavy top-mass expansion \cite{Harlander:2009bw, Pak:2009bx, Harlander:2009mq, Pak:2009dg} and leading contributions to the top/bottom interference \cite{Mueller:2015lrx,Lindert:2017pky} are known. Beyond inclusive production, the only available NNLO result is the production of a Higgs boson in association with a jet in the infinite top-mass limit~\cite{Boughezal:2015dra,Boughezal:2015aha,Chen:2016zka}, while cross sections for $H+n$-jets, $n=2,3$, are known only at NLO in the heavy top-mass expansion~\cite{Campbell:2006xx,Cullen:2013saa}. 

In the SMEFT, most studies have been performed at LO, typically using approximate rescaling factors obtained from SM calculations. Higher-order results have only been considered when existing SM calculations could be readily used within the SMEFT. The simplest examples are the inclusion of higher orders in the strong coupling to the contribution of two specific dimension-six operators, namely the Yukawa operator $(\Phi^\dagger \Phi)  \bar Q_L  \Phi  q_R$ and the gluon-fusion operator $(\Phi^\dagger \Phi) GG$. The former can be accounted for by a straightforward modification of the Yukawa coupling of the corresponding heavy quark, $b$ or $t$, while the latter involves the computation of  contributions identical to SM calculations in the limit of an infinitely-heavy top quark.  Results for the inclusive production cross section including modified top and bottom Yukawa couplings and an additional direct Higgs-gluons interaction are available at NNLO~\cite{Brooijmans:2016vro} and at N$^3$LO~\cite{Harlander:2016hcx, Anastasiou:2016hlm}. 
At the differential level, phenomenological studies at LO have shown the relevance of the high transverse momentum region of the Higgs boson in order to resolve degeneracies among operators present at the inclusive level~\cite{Grojean:2013nya, Schlaffer:2014osa, Buschmann:2014twa, Maltoni:2016yxb}. Recently,  the calculation of the Higgs spectrum at NLO+NNLL level for the Yukawa (both $b$ and $t$) and Higgs-gluons operator has appeared~\cite{Grazzini:2016paz,Grazzini:2017szg}. 

The purpose of this work is to provide the contribution of the chromomagnetic operator $ \bar Q_L \Phi  \sigma q_R G$  to inclusive Higgs production at NLO in QCD, thereby completing the set of predictions (involving only $CP$-even interactions) needed to achieve NLO accuracy in the SMEFT for this process. The first correct computation at one-loop of the contribution of chromomagnetic operator of the top quark to $gg\to H$  has appeared in the erratum of ref.~\cite{Degrande:2012gr} and later confirmed in refs.~\cite{Maltoni:2016yxb, Grazzini:2017szg}. The LO contribution of the chromomagnetic operator of the top-quark to $H+$jet was computed in ref.~\cite{Maltoni:2016yxb}. An important conclusion drawn in ref.~\cite{Maltoni:2016yxb} was that even when the most stringent (and still approximate) constraints from $t\bar t$ production are considered~\cite{Franzosi:2015osa}, this operator sizably affects Higgs production, both in gluon fusion (single and double Higgs) and $t\bar t H$ production. 

At LO the chromomagnetic operator enters Higgs production in gluon fusion at one loop.  Therefore NLO corrections in QCD entail two-loop virtual and one-loop real contributions. The latter can nowadays easily be computed using an automated approach. The former, however, involve a non-trivial two-loop computation that requires analytic multi-loop techniques and a careful treatment of the renormalisation and mixing in the SMEFT, both of which are presented in this work for the first time. In particular, while the full mixing pattern of the SMEFT at one loop is known~\cite{Jenkins:2013zja,Jenkins:2013wua,Alonso:2013hga}, a new two-loop counterterm enters our computation, and we provide its value for the first time here. Moreover, we present very compact analytic results for all the relevant amplitudes up to two loop order. Focusing on possibly anomalous contributions in top-quark-Higgs interactions, we then consider the phenomenological impact of the NLO corrections, including also the Yukawa operator and the gluon-fusion operator at NLO by implementing the respective virtual two-loop matrix elements into the {\sc MadGraph5\_aMC@NLO}  framework \cite{Alwall:2014hca}. This allows us to compute total cross sections as well as to perform event generation at NLO plus parton shower (NLO+PS) that can be directly employed in experimental analyses.    

The paper is organised as follows. In section~\ref{sec:setup} we establish our notations and set up the calculation by identifying the terms in the perturbative expansion that are unknown and need to be calculated. In section~\ref{sec:virtuals} we describe in detail the computation of the two-loop virtual contributions and the renormalisation procedure and we provide compact analytic expressions for the finite parts of the two-loop amplitudes. We also briefly discuss the leading logarithmic renormalisation group running of the Wilson coefficients. In section~\ref{sec:pheno} we perform a phenomenological study at NLO, in particular of the behaviour of the QCD and EFT expansion at the total inclusive level and provide predictions for the $p_T$ spectrum of the Higgs via a NLO+PS approach.


\section{Gluon fusion in the SM Effective Field Theory}
\label{sec:setup}
\begin{figure}[t]
\centering
\includegraphics[width=.99\linewidth]{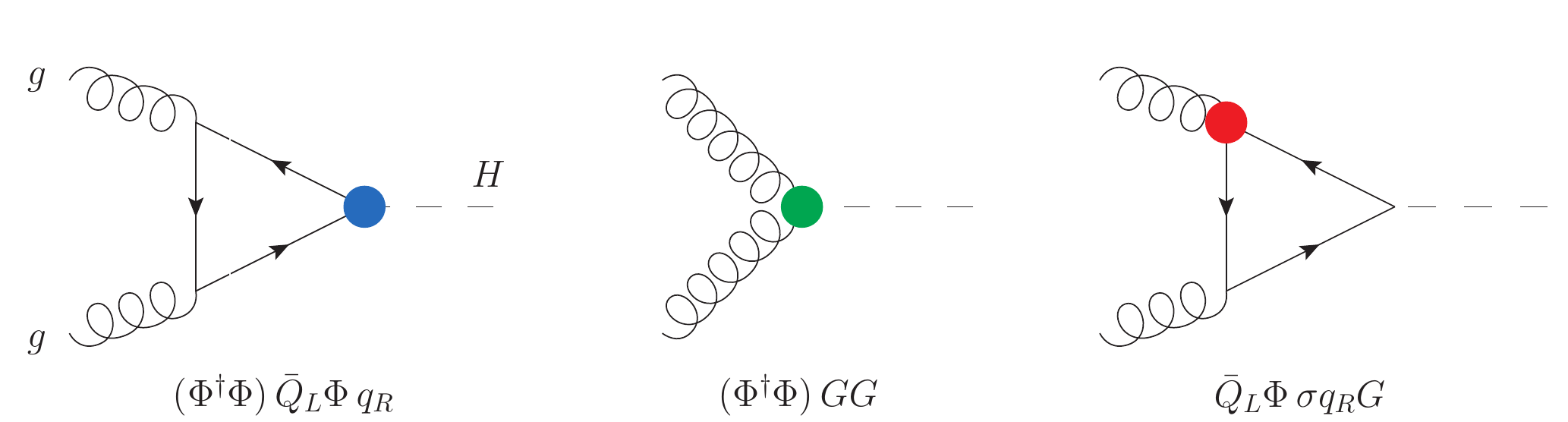}
\caption{\label{fig:gglo} Representative diagrams contributing to gluon-fusion amplitudes with one insertion of the three relevant operators. Heavy quarks, $b$ or $t$, provide the leading contributions to the first and third amplitudes. Note that for chromomagnetic operator, $ \bar Q_L \Phi  \sigma q_R G$, a diagram featuring the four point gluon-quark-quark-Higgs interaction is also present (not shown).
 }
\end{figure}

The goal of this paper is to study the production of a Higgs boson in hadron collisions in the SMEFT, {i.e.}, the SM supplemented by a complete set of operators of dimension six,
\beq\label{eq:L_EFT}
\cL_{EFT}  = \cL_{SM} + \sum_{i} \left(\frac{C_i^b}{\Lambda^2}\,\cO_i + \textrm{h.c.}\right) \,.
\eeq
The sum in eq.~\eqref{eq:L_EFT} runs over a basis of operators  $\cO_i$ of dimension six, $\Lambda$ is the scale of new physics  and $C_i^b$ are the (bare) Wilson coefficients, multiplying the effective operators. A complete and independent set of operators of dimension six is known~\cite{Grzadkowski:2010es,Buchmuller:1985jz}. In this paper, we are only interested in those operators that modify the contribution of the heavy quarks, bottom and top quarks, to Higgs production in gluon fusion. Focusing on the top quark, there are three operators of dimension six that contribute to the gluon-fusion process,
\begin{align}
\label{eq:O1}
\cO_{1} &\,=\left(\Phi^\dagger\Phi-\frac{v^2}{2}\right)\,\,\overline{Q}_{L} \tilde{\Phi}\, t_{R} \,,\\
\label{eq:O2}
\cO_{2} &\,= \,g_s^2\,\left(\Phi^\dagger\Phi-\frac{v^2}{2}\right)\,G_{\mu\nu}^aG^{\mu\nu}_a\,,\\
\label{eq:O3}
\cO_{3} &\,= g_s\,\overline{Q}_{L} \tilde{\Phi} \, T^a \,\sigma^{\mu\nu}  t_{R}\,G_{\mu\nu}^a \,,
\end{align}
where $g_s$ is the (bare) strong coupling constant and $v$ denotes the vacuum expectation value (vev) of the Higgs field $\Phi$ ($\tilde \Phi=i\sigma_2 \Phi$). $Q_{L}$ is the left-handed quark $SU(2)$-doublet containing the top quark, $t_R$ is the right-handed $SU(2)$-singlet top quark, and $G_{\mu\nu}^a$ is the gluon field strength tensor. Finally, $T^a$ is the generator of the fundamental representation of $SU(3)$ (with $[T^a,T^b]=\frac12 \delta^{ab}$) and $\sigma^{\mu\nu}=\frac{i}{2}[\gamma^\mu,\gamma^\nu]$, with $\gamma^\mu$ the Dirac gamma matrices.
Two comments are in order.  First, the corresponding operators $\cO_{1}$ and $\cO_{3}$ for the $b$ quark can be obtained by simply making the substitutions \{$\tilde \Phi \to \Phi$, $t_{R} \to b_{R}$\}. Second, while $\cO_{2}$ is hermitian  $\cO_{1}$ and $\cO_{3}$ are not.\footnote{Note that in eq.~(\ref{eq:L_EFT}) we adopt the convention to include the hermitian conjugate for all operators, be they hermitian or not. This means that the overall contribution from $\cO_{2}$ in $\cL_{EFT}$  is actually $ 2 C_2 \cO_{2}/\Lambda^2$.} In this work, we focus on the $CP$-even contributions of $\cO_{1}$ and $\cO_{3}$. For this reason, all the Wilson coefficients $C_i$ with $i=1,2,3$ are real.
Representative Feynman diagrams contributing at LO are shown in fig.~\ref{fig:gglo}.

In the SM and at leading order (LO) in the strong coupling the gluon-fusion process is mediated only by quark loops. This contribution is proportional to the mass of the corresponding quark and therefore heavy quarks dominate. While we comment on the $b$ (and possibly $c$) contributions later, let us focus on the leading contributions coming from the top quark, {i.e.}, the contributions from the operators of dimension six shown in eqs.~(\ref{eq:O1} -~\ref{eq:O3}). The (unrenormalised) amplitude can be cast in the form
\begin{align}\label{eq:amp_LO}
\cA_b&(g\,g\to H) = \frac{i\,S_{\eps}\,\mu^{-2\eps}\,\alpha_s^b}{\pi}\,\left[(p_1\cdot p_2)\,(\eps_1\cdot \eps_2)-(p_1\cdot \eps_2)\,(p_2\cdot \eps_1)\right]\,\left[\frac{1}{v}\,\cA_{b,0}(m_t^b,m_H)\right.\\
\nonumber&\,\left.+\frac{C_{1}^b\,v^2}{\sqrt{2}\,\Lambda^2}\,\cA_{b,1}(m_t^b,m_H)+\frac{C_{2}^b\,v}{\Lambda^2}\,\cA_{b,2}(m_t^b,m_H)
+\frac{C_{3}^b}{\sqrt{2}\,\Lambda^2}\,\cA_{b,3}(m_t^b,m_H)\right]+\ord(1/\Lambda^4)\,,
\end{align}
where $\alpha_s^b = g_s^2/(4\pi)$ denotes the bare QCD coupling constant and $m_H$ and $m_t^b$ are the bare masses of the Higgs boson and the top quark. The factor $S_\eps = e^{-\gamma_E\eps}\,(4\pi)^{\eps}$ is the usual $\overline{\textrm{MS}}$ factor, with $\gamma_E=-\Gamma'(1)$ the Euler-Mascheroni constant and $\mu$ is the  scale introduced by dimensional regularisation.  For $i=0$, the form factor $\cA_{b,i}$ denotes the unrenormalised SM contribution to gluon fusion~\cite{Georgi:1977gs}, while for $i>0$ it denotes the form factor with a single\footnote{According to our power counting rules, multiple insertions of an operator of dimension six correspond to contributions of $\ord(1/\Lambda^4)$ in the EFT, and so they are neglected.} operator $\cO_i$ inserted~\cite{Choudhury:2012np,Degrande:2012gr,Grazzini:2016paz}. The normalisation of the amplitudes is chosen such that all coupling constants, as well as all powers of the vev $v$, are explicitly factored out. Each form factor admits a perturbative expansion in the strong coupling,
\beq
\cA_{b,i}(m_t^b,m_H) = \sum_{k=0}^\infty\left(\frac{S_\eps\,\mu^{-2\eps}\,\alpha_s^b}{\pi}\right)^k\,\cA_{b,i}^{(k)}(m_t^b,m_H)\,.
\eeq
 Some comments  about these amplitudes are in order. First, after electroweak symmetry breaking, the operator $\cO_1$ only amounts to a rescaling of the Yukawa coupling, {i.e.}, $\cA_{b,1}$ is simply proportional to the bare SM amplitude. Second, at LO the operator $\cO_2$ contributes at tree level, while the SM amplitude and the contributions from $\cO_1$ and $\cO_3$ are loop-induced. Finally, this process has the unusual feature that the amplitude involving the chromomagnetic operator $\cO_3$ is ultraviolet (UV) divergent, and thus requires renormalisation, already at LO~\cite{Degrande:2012gr,Maltoni:2016yxb,Grazzini:2017szg}. The UV divergence is absorbed into the effective coupling that multiplies the operator $\cO_2$, which only enters at tree level at LO. The renormalisation at NLO will be discussed in detail in section~\ref{sec:virtuals}.

The goal of this paper is to compute the NLO corrections to the gluon-fusion process with an insertion of one of the dimension six operators in eqs.~(\ref{eq:O1} -~\ref{eq:O3}). We emphasise that a complete NLO computation requires one to consider the set of all three operators in eq.~(\ref{eq:O1} -~\ref{eq:O3}), because they mix under renormalisation~\cite{Jenkins:2013zja,Jenkins:2013wua,Alonso:2013hga}. At NLO, we need to consider both virtual corrections to the LO process $g\,g\to H$ as well as real corrections due to the emission of an additional parton in the final state. Starting from NLO, also partonic channels with a quark in the initial state contribute. Since the contribution from $\cO_1$ is proportional to the SM amplitude, the corresponding NLO corrections can be obtained from the NLO corrections to gluon-fusion in the SM including the full top-mass dependence~\cite{Graudenz:1992pv,Spira:1995rr,Anastasiou:2006hc,Anastasiou:2009kn}. The NLO contributions from $\cO_2$ are also known, because they are proportional to the NLO corrections to gluon-fusion in the SM in the limit where the top quark is infinitely heavy~\cite{Dawson:1990zj} (without the higher-order corrections to the matching coefficient). In particular, the virtual corrections to the insertion of $\cO_2$ are related to the QCD form factor, which is known through three loops in the strong coupling~\cite{Kramer:1986sg,Matsuura:1987wt,Matsuura:1988sm,Harlander:2000mg,Moch:2005tm,Moch:2005id,Gehrmann:2005pd,Gehrmann:2010ue,Gehrmann:2010tu,Baikov:2009bg}. Hence, the only missing ingredient is the NLO contributions to the process where the chromomagnetic operator $\cO_3$ is inserted. The computation of this ingredient, which is one of the main results of this paper, will be presented in detail in the next section.

As a final comment, we note that starting at two loops other operators of EW and QCD nature will  affect $gg\to H$. In the case of EW interactions, by just looking at the SM EW contributions~\cite{Degrassi:2004mx,Actis:2008ug},  it is easy to see that many operators featuring the Higgs field will enter, which in a few cases could also lead to constraints, see, {e.g.}, the trilinear Higgs self coupling ~\cite{Gorbahn:2016uoy,Degrassi:2016wml}. In the case of QCD interactions, operators not featuring the Higgs field will enter, which, in general, can be more efficiently bounded from other observables. For example, the operator $g_s f^{abc} G_{a\mu}^{\nu} G_{b\nu}^{\lambda} G_{c\lambda}^{\mu}$
contributes at two loops in $gg\to H$ and at one loop in $gg\to Hg$. The latter process has been considered in ref.~\cite{Ghosh:2014wxa}, where effects on the transverse momentum of the Higgs were studied. For the sake of completeness, we have reproduced these results in our framework, and by considering the recent constraints on this operator from multi-jet observables~\cite{Krauss:2016ely}, we have confirmed that the Higgs $p_T$ cannot be significantly affected. For this reason we do not discuss further this operator in this paper. Four-fermion operators also contribute starting at two loops to gluon fusion but as these modify observables related to top quark physics at leading order~\cite{Buckley:2015lku,Zhang:2017mls} we expect them to be independently constrained and work under the assumption that they cannot significantly affect gluon fusion.


\section{Virtual corrections}
\label{sec:virtuals}

\subsection{Computation of the two-loop amplitudes}

In this section we describe the virtual corrections to the LO amplitudes in eq.~\eqref{eq:amp_LO}. For the sake of the presentation we focus here on the calculation involving a top quark and discuss later on how to obtain the corresponding results for the bottom quark.  With the exception of the contributions from $\cO_2$, all processes are loop-induced, and so the virtual corrections require the computation of two-loop form factor integrals with a closed heavy-quark loop and two external gluons. We have implemented the operators in eqs.~(\ref{eq:O1} -~\ref{eq:O3}) into QGraf~\cite{Nogueira:1991ex}, and we use the latter to generate all the relevant Feynman diagrams. The QGraf output is translated into {\tt FORM}~\cite{Vermaseren:2000nd,Kuipers:2012rf} and Mathematica using a custom-made code. The tensor structure of the amplitude is fixed by gauge-invariance to all loop orders, cf. eq.~\eqref{eq:amp_LO}, and we can simply project each Feynman diagram onto the transverse polarisation tensor. The resulting scalar amplitudes are then classified into distinct integral topologies, which are reduced to master integrals using {\tt FIRE} and {\tt LiteRed}~\cite{Smirnov:2008iw,Lee:2012cn,Smirnov:2013dia,Lee:2013mka,Smirnov:2014hma}. After reduction, we can express all LO and NLO amplitudes as a linear combination of one and two-loop master integrals.

The complete set of one- and two-loop master integrals is available in the literature~\cite{Bonciani:2003te,Bonciani:2003hc,Aglietti:2006tp,Anastasiou:2006hc} in terms of harmonic polylogarithms (HPLs)~\cite{Remiddi:1999ew},
\beq
H(a_1,\ldots,a_w;z) = \int_0^zdt\,f(a_1,t)\,H(a_2,\ldots,a_w;z)\,,
\eeq
with
\beq
f(1,t) = \frac{1}{1-t}\,,\qquad f(0,t) = \frac{1}{t}\,,\qquad f(-1,t) = \frac{1}{1+t}\,.
\eeq
In the case where all the $a_i$'s are zero, we define,
\beq
H(\underbrace{0,\ldots,0}_{w\textrm{ times}};z) = \frac{1}{w!}\,\log^wz\,.
\eeq
The number of integrations $w$ is called the \emph{weight} of the HPL.
The only non-trivial functional dependence of the master integrals is through the ratio of the Higgs and the top masses, and it is useful to introduce the following variable,
\beq\label{eq:x_def}
\tau = \frac{m_H^2}{m_t^2} = -\frac{(1-x)^2}{x}\,,
\eeq
or equivalently
\beq\label{eq:x_def_sqrt}
x = \frac{\sqrt{1-4/\tau}-1}{\sqrt{1-4/\tau}+1}\,.
\eeq
The change of variables in eq.~\eqref{eq:x_def} has the advantage that the master integrals can be written as a linear combination of HPLs in $x$.
In the kinematic range that we are interested in, $0 <m_H^2 <4m_t^2$, the variable $x$ is a unimodular complex number, $|x|=1$, and so it can be conveniently parametrised in this kinematics range by an angle $\theta$,
\beq\label{eq:theta_def}
x=e^{i\theta}\,,\qquad 0<\theta<\pi\,.
\eeq
In terms of this angle, the master integrals can be expressed in terms of (generalisations of) Clausen functions (cf. ref.~\cite{Davydychev:2000na,Davydychev:2003mv,Kalmykov:2005hb,Kalmykov:2004xg,Anastasiou:2006hc} and references therein),
\beq\label{eq:clausen}
\cl_{m_1,\ldots,m_k}(\theta) = \left\{\begin{array}{ll}
\displaystyle \textrm{Re}\, H_{m_1,\ldots, m_k}\big(e^{i\theta}\big)\,, &\textrm{ if $k+w$ even}\,,\\
\displaystyle \textrm{Im}\, H_{m_1,\ldots, m_k}\big(e^{i\theta}\big)\,, &\textrm{ if $k+w$ odd}\,,
\end{array}\right.
\eeq
where we used the notation
\beq
H_{m_1,\ldots, m_k}(z) = H(\!\!\!\!\!\underbrace{0,\ldots,0}_{(|m_1|-1)\textrm{ times}}\!\!\!\!\!,\sigma_1,\ldots,\!\!\!\!\!\underbrace{0,\ldots,0}_{(|m_k|-1)\textrm{ times}}\!\!\!\!\!,\sigma_k;z)\,,\qquad \sigma_i\equiv \textrm{sign}(m_i)\,.
\eeq
The number $k$ of non-zero indices is called the \emph{depth} of the HPL.

Inserting the analytic expressions for the master integrals into the amplitudes, we can express each amplitude as a Laurent expansion in $
\eps$ whose coefficients are linear combinations of the special functions we have just described. The amplitudes have poles in $\eps$ which are of both ultraviolet (UV) and infrared (IR) nature, whose structure is discussed in the next section.


\subsection{UV \& IR pole structure}
In this section we discuss the UV renormalisation and the IR pole structure of the LO and NLO amplitudes. We start by discussing the UV singularities. We work in the $\overline{\textrm{MS}}$ scheme, and we write the bare amplitudes as a function of the renormalised amplitudes as,
\beq
\cA_b(\alpha_s^b,C^b_i,m_t^b,m_H) = Z_g^{-1}\,\cA(\alpha_s(\mu^2),C_i(\mu^2),m_t(\mu^2),m_H,\mu)\,,
\eeq
where $Z_g$ is the field renormalisation constant of the gluon field and $\alpha_s(\mu^2)$, $C_i(\mu^2)$ and $m_t(\mu^2)$ are the renormalised strong coupling constant, Wilson coefficients and top mass in the $\overline{\textrm{MS}}$ scheme, and $\mu$ denotes the renormalisation scale. The renormalised parameters are related to their bare analogues through
\beq\bsp
S_{\eps}\,\alpha_s^b &\,=\mu^{2\eps}\,Z_{\alpha_s}\,\alpha_s(\mu^2)\,,\\
C_i^b &\,= \mu^{a_i\eps}\,Z_{C,ij}\,C_j(\mu^2)\,,\\
m_t^b &\,= m_t(\mu^2) + \delta m_t\,,
\esp\eeq
with $(a_1,a_2,a_3) = (3,0,1)$. Unless stated otherwise, all renormalised quantities are assumed to be evaluated at the arbitrary scale $\mu^2$ throughout this section. We can decompose the renormalised amplitude into the contributions from the SM and the effective operators, similar to the decomposition of the bare amplitude in eq.~\eqref{eq:amp_LO}
\beq\bsp
\cA&(g\,g\to H) = \frac{i\,\alpha_s}{\pi}\,\left[(p_1\cdot p_2)\,(\eps_1\cdot \eps_2)-(p_1\cdot \eps_2)\,(p_2\cdot \eps_1)\right]\,\left[\frac{1}{v}\,\cA_{0}(m_t,m_H)\right.\\
&\,\left.+\frac{C_{1}\,v^2}{\sqrt{2}\,\Lambda^2}\,\cA_{1}(m_t,m_H)+\frac{C_{2}\,v}{\Lambda^2}\,\cA_{2}(m_t,m_H)
+\frac{C_{3}}{\sqrt{2}\,\Lambda^2}\,\cA_{3}(m_t,m_H)\right] + \ord(1/\Lambda^4)\,,
\esp\eeq
and each renormalised amplitude admits a perturbative expansion in the renormalised strong coupling constant,
\beq
\cA_{i}(m_t,m_H) = \sum_{k=0}^\infty\left(\frac{\alpha_s}{\pi}\right)^k\,\cA_{i}^{(k)}(m_t,m_H)\,.
\eeq

The presence of the effective operators alters the renormalisation of the SM parameters. Throughout this section we closely follow the approach of ref.~\cite{Maltoni:2016yxb},  where the renormalisation of the operators at one loop was described.
 The one-loop UV counterterms for the strong coupling constant and the gluon field are given by
\beq\bsp
Z_g &\,= 1+\delta Z_{g,SM} + \frac{\alpha_s}{\pi}
\,\frac{C_3}{\Lambda^2}\,\frac{1}{\eps}\,\left(\frac{\mu^2}{m_t^2}\right)^{\eps}{\sqrt{2}\,v}\,m_t+\ord(\alpha_s^2)\,,\\
Z_{\alpha_s} &\,= 1+\delta Z_{\alpha_s,SM}-\frac{\alpha_s}{\pi}
\,\frac{C_3}{\Lambda^2}\,\frac{1}{\eps}\,\left(\frac{\mu^2}{m_t^2}\right)^{\eps}{\sqrt{2}\,v}\,m_t+\ord(\alpha_s^2)\,,
\esp\eeq
where $\delta Z_{g,SM}$ and $\delta Z_{\alpha_s,SM}$ denote the one-loop UV counterterms in the SM,
\beq\bsp
\delta Z_{g,SM} &\,=  \frac{\alpha_s}{\pi}\,\frac{1}{6\eps}\,\left(\frac{\mu^2}{m_t^2}\right)^{\eps}+\ord(\alpha_s^2)\,,\\
\delta Z_{\alpha_s,SM} &\,=  -\frac{\alpha_s}{4\pi}\,\frac{\beta_0}{\eps}-\frac{\alpha_s}{\pi}\,\frac{1}{6\eps}\,\left(\frac{\mu^2}{m_t^2}\right)^{\eps}+\ord(\alpha_s^2)\,,
\esp\eeq
and $\beta_{0}$ is the one-loop QCD $\beta$ function,
\beq
\beta_0 = \frac{11N_c}{3}-\frac{2}{3}N_f\,,
\label{eq:beta0}
\eeq
where $N_c=3$ is the number of colours and $N_f=5$ is the number of massless flavours. We work in a decoupling scheme and we include a factor $\left({\mu^2}/{m_t^2}\right)^{\eps}$ into the counterterm. As a result only massless flavours contribute to the running of the strong coupling, while the top quark effectively decouples~\cite{Dawson:1990zj}.
The renormalisation of the strong coupling and the gluon field are modified by the presence of the dimension six operators, but the effects cancel each other out~\cite{Degrande:2012gr}.  Similarly, the renormalisation of the top mass is modified by the presence of the effective operators,
\beq\label{eq:mt_renorm}
\delta m_t = \delta m_{t}^{SM} -\frac{\alpha_s}{\pi}\,\frac{C_3}{\Lambda^2}\,\frac{1}{\eps}\,\left(\frac{\mu^2}{m_t^2}\right)^{\eps} 2\sqrt{2}\,v\,m_t^2+\ord(\alpha_s^2)\,,
\eeq
where the SM contribution is
\beq
\delta m_{t}^{SM} = -\frac{\alpha_s}{\pi}\,\frac{m_t}{\eps}+\ord(\alpha_s^2)\,.
\eeq
In eq.~\eqref{eq:mt_renorm} we again include the factor $\left({\mu^2}/{m_t^2}\right)^{\eps}$ into the counterterm in order to decouple the effects from operators of dimension six from the running of the top mass in the $\overline{\textrm{MS}}$ scheme.

The renormalisation of the effective couplings $C_i^b$ is more involved, because the operators in eqs.~(\ref{eq:O1} -~\ref{eq:O3}) mix under renormalisation. The matrix $Z_{C}$ of counterterms can be written in the form
\beq\label{eq:Z_C_def}
Z_{C} = \mathbb{1} + \delta Z_{C}^{(0)} + \frac{\alpha_s}{\pi}\,\delta Z_{C}^{(1)} +\ord(\alpha_s^2)\,.
\eeq
We have already mentioned that the amplitude $\cA_{b,3}$ requires renormalisation at LO in the strong coupling, and the UV divergence is proportional to the LO amplitude $\cA_{b,2}^{(0)}$~\cite{Degrande:2012gr,Maltoni:2016yxb,Grazzini:2017szg}. As a consequence, $\delta Z_{C}^{(0)}$ is non-trivial at LO in the strong coupling,
\beq
\delta Z_{C}^{(0)} = \left(
\begin{array}{ccc}
 0 & 0 & 0 \\
 0 & 0 & \dfrac{\sqrt{2} \,{m_t}}{16\pi^2\,\epsilon \, v} \\
 0 & 0 & 0 \\
\end{array}
\right)\,.
\eeq
At NLO, we also need the contribution $\delta Z_{C}^{(1)}$ to eq.~\eqref{eq:Z_C_def}. We have
\beq
\delta Z_{C}^{(1)} = \left(
\begin{array}{ccc}
 -\dfrac{1}{\epsilon } & 0 & \dfrac{8
   {m_t}^2}{\epsilon\,  v^2} \\
 0 & 0 & z_{23} \\
 0 & 0 & \dfrac{1}{6\, \epsilon } \\
\end{array}
\right)\,,
\eeq
where, apart from  $z_{23}$,  all the entries are known~\cite{Jenkins:2013zja,Jenkins:2013wua,Alonso:2013hga}.  $z_{23}$ corresponds to the counterterm that absorbs the two-loop UV divergence of the operator $\cO_3$, which is proportional to the tree-level amplitude $\cA_{b,2}^{(0)}$ in our case. This counterterm is not available in the literature, yet we can extract it from our computation. NLO amplitudes have both UV and IR poles, and so we need to disentangle the two types of divergences if we want to isolate the counterterm $z_{23}$. We therefore first review the structure of the IR divergences of NLO amplitudes, and we will return to the determination of the counterterm $z_{23}$ at the end of this section.

A one-loop amplitude with massless gauge bosons has IR divergences, arising from regions in the loop integration where the loop momentum is soft or collinear to an external massless leg. The structure of the IR divergences is universal in the sense that it factorises from the underlying hard scattering process. More precisely, if $\cA^{(1)}$ denotes a renormalised one-loop amplitude describing the production of a colourless state from the scattering of two massless gauge bosons, then we can write~\cite{Catani:1996vz}
\beq\label{eq:IR_structure}
\cA^{(1)} = {\bf I}^{(1)}(\eps)\,\cA^{(0)} + \cR\,,
\eeq
where $\cA^{(0)}$ is the tree-level amplitude for the process and $\cR$ is a process-dependent remainder that is finite in the limit $\eps\to 0$. The quantity ${\bf I}^{(1)}(\eps)$ is universal (in the sense that it does not depend on the details of the hard scattering) and is given by
\beq
{\bf I}^{(1)}(\eps) = -\frac{e^{-\gamma_E\eps}}{\Gamma(1-\eps)}\,\left(\frac{-s_{12}-i0}{\mu^2}\right)^{-\eps}\,\left(\frac{3}{\eps^2}+\frac{\beta_0}{2\eps}\right)\,,
\eeq
where $s_{12} = 2p_1p_2$ denotes the center-of-mass energy squared of the incoming gluons.

Since in our case most amplitudes are at one loop already at LO, we have to deal with two-loop amplitudes at NLO. However, since the structure of the IR singularities is independent of the details of the underlying hard scattering, eq.~\eqref{eq:IR_structure} remains valid for two-loop amplitudes describing loop-induced processes, and we can write
\beq\label{eq:IR_structure_i}
\cA^{(1)}_i = {\bf I}^{(1)}(\eps)\,\cA^{(0)}_i + \cR_i\,, \qquad 0\le i\le 3\,.
\eeq
We have checked that our results for amplitudes which do not involve the operator $\cO_3$ have the correct IR pole structure at NLO. For $\cA^{(1)}_3$, instead, we can use eq.~\eqref{eq:IR_structure_i} as a constraint on the singularities of the amplitude. This allows us to extract the two-loop UV counterterm $z_{23}$. We find
\beq\label{eq:z23}
z_{23} = \frac{{m_t}}{16\,\pi^2\,v\,\sqrt{2}}\,\left(-\frac{5}{6\,\epsilon^2}+\frac{23}{4\,\epsilon}\right)\,.
   \eeq
   Note that the coefficient of the double pole is in fact fixed by requiring the anomalous dimension of the effective couplings to be finite. We have checked that eq.~\eqref{eq:z23} satisfies this criterion, which is a strong consistency check on our computation.

 Let us conclude our discussion of the renormalisation with a comment on the relationship between the renormalised amplitudes in the SM and the insertion of the operator $\cO_1$. We know that the corresponding unrenormalised amplitudes are related by a simple rescaling, and the constant of proportionality is proportional to the ratio $C_1^b/m_t^b$. There is a priori no reason why such a simple relationship should be preserved by the renormalisation procedure. In (the variant of) the $\overline{\textrm{MS}}$-scheme that we use, the renormalised amplitudes are still related by this simple scaling. This can be traced back to the fact that the $\overline{\textrm{MS}}$ counterterms are related by
 \beq
 \delta m_t^{SM} = \frac{\alpha_s}{\pi}\,\big(Z_C^{(1)}\big)_{11} + \ord(\alpha_s^2)\,.
 \eeq
 If the top mass and the Wilson coefficient $C_1^b$ are renormalised using a different scheme which breaks this relation between the counterterms, the simple relation between the amplitudes $\cA_0^{(1)}$ and $\cA_1^{(1)}$ will in general not hold after renormalisation.

\subsection{Analytic results for the two-loop amplitudes}
In this section we present the analytic results for the renormalised amplitudes that enter the computation of the gluon-fusion cross section at NLO with the operators in eqs.~(\ref{eq:O1} -- \ref{eq:O3}) included. We show explicitly the one-loop amplitudes up to $\mathcal{O}(\eps^2)$ in dimensional regularisation, as well as the finite two-loop remainders $\cR_i$ defined in eq.~\eqref{eq:IR_structure}.
The amplitudes have been renormalised using the scheme described in the previous section and all scales are fixed to the mass of the Higgs boson, $\mu^2=m_H^2$.

The operator $\cO_2$ only contributes at one loop at NLO, and agrees (up to normalisation) with the one-loop corrections to Higgs production via gluon-fusion~\cite{Dawson:1990zj}. The amplitude is independent of the top mass through one loop, and so it evaluates to a pure number,
\beq
\cA_2^{(0)} = -32\,\sqrt{2}\,\pi^2 {\rm~~and~~} \cR_2 = 16\,i\pi^3\,\beta_0\,,
\eeq
where $\beta_0$ is defined in eq.~\eqref{eq:beta0}. The remaining amplitudes have a non-trivial functional dependence on the top mass through the variables $\tau$ and $\theta$ defined in eq.~\eqref{eq:x_def} and ~\eqref{eq:theta_def}.
We have argued in the previous section that in the $\overline{\textrm{MS}}$-scheme the renormalised amplitudes $\cA_0^{(1)}$ and $\cA_1^{(1)}$ are related by a simple rescaling,
\beq
\cA_1^{(1)} = -\frac{1}{m_t}\,\cA_0^{(1)}\,.
\eeq
We therefore only present results for the SM contribution and the contribution from $\cO_3$. We have checked that our result for the two-loop amplitude in the SM agrees with the results of ref.~\cite{Graudenz:1992pv,Spira:1995rr,Anastasiou:2006hc,Anastasiou:2009kn}. The two-loop amplitude $\cA_3^{(1)}$ is genuinely new and is presented here for the time.

The one-loop amplitude in the SM can be cast in the form
\beq
\cA^{(0)}_0 = a_0 +\eps\left(a_1 + \log\tau\,a_0\right) + \eps^2 \left(a_2 + \log\tau\,a_1 + \frac{1}{2}\log^2\tau\,a_0\right)+\ord(\eps^3)\,,
\eeq
where the coefficients $a_i$ are given by
\begin{align}
a_0 &\,= \frac{2 \theta ^2}{\tau ^2}-\frac{\theta ^2+4}{2 \tau }\,,\\
\nonumber a_1 &\,= \frac{1}{\tau }\,\left(1-\frac{4}{\tau }\right)\, \left[4\, \text{Cl}_{-3}(\theta )+2\, \theta\,  \text{Cl}_{-2}(\theta )-3 \zeta_3\right]+\frac{2 \theta ^2}{\tau ^2}-\frac{6}{\tau } +\frac{2 \theta}{\sqrt{(4-\tau ) \tau }}\left(1-\frac{4}{\tau }\right)\,,\\
\nonumber a_2 &\,=
\frac{1}{\tau }\left(1-\frac{4}{\tau }\right) \left[2\, \text{Cl}_{-2}(\theta )^2-4\, \theta\,  \text{Cl}_{-2,-1}(\theta )-\frac{\theta ^4}{6}-\frac{\pi ^2 \theta ^2}{24}\right]+\frac{2}{\tau ^2} [\theta ^2+6\, \zeta_3-4\, \theta\,  \text{Cl}_{-2}(\theta )\\
\nonumber&\,-8\, \text{Cl}_{-3}(\theta )]-\frac{1}{\tau }\left(14+\frac{\pi ^2}{6}\right)
-\frac{2}{\sqrt{(4-\tau ) \tau }} \left(1-\frac{4}{\tau }\right) [\theta\,  \log(4-\tau)-2\, \text{Cl}_{-2}(\theta )-3\, \theta] \,.
\end{align}
The finite remainder of the two-loop SM amplitude is
\begin{align}
\nonumber\cR_0&\, = -\frac{i \pi \,\beta _0 }{16 \tau }\,\left(\theta ^2 \tau -4 \theta ^2+4 \tau \right)-\frac{4}{\tau }\, \left(1-\frac{4}{\tau }\right) \,\Bigg[3\, \theta\,  \text{Cl}_{1,-2}(\theta )+6\, \text{Cl}_{1,-3}(\theta )+3\, \text{Cl}_{2,-2}(\theta )\\
\nonumber&\,\qquad-\frac{4}{3}\, \theta\,  \text{Cl}_{-2}(\theta )+\frac{17}{4}\, \theta \, \text{Cl}_2(\theta )-\frac{8}{3}\, \text{Cl}_{-3}(\theta )+\frac{55 }{12}\,\text{Cl}_3(\theta )-\frac{3 }{8}\,\zeta_2\, \theta ^2\\
&\, \qquad+\frac{9}{4}\, \zeta _3\, \log\tau-\frac{31 }{12}\,\zeta _3+\frac{51}{16}\,\zeta _4-\frac{5 }{64}\,\theta^4+\frac{59}{48}\, \theta^2\, \log\tau+\frac{25}{48}\,\theta ^2+\log\tau+\frac{21}{4}\Bigg]\\
\nonumber&\,{+ \frac{1}{3 \tau ^2}}\left[-28\, \theta\,  \text{Cl}_2(\theta )-28 \,\text{Cl}_3(\theta )+28\, \zeta _3+5\, \theta ^2\, \log\tau +28\, \theta ^2-48\, \log\tau-252\right]\\
\nonumber&-\frac{4}{\tau ^2 \sqrt{(4-\tau ) \tau }}\,\left(1-\frac{3 \tau }{4}+\frac{\tau ^2}{8}\right)\,R(\theta) +\frac{\theta\,  (4-\tau ) }{12\, \tau  \sqrt{(4-\tau ) \tau }}\,\left(13 \theta ^2+24 \log\tau -16\right)\,,
\end{align}
where we have defined the function
\beq\bsp
R(\theta)&\, = -\frac{16}{3}\, \theta^2\, \text{Cl}_{-2}(\theta)+\frac{28}{3}\, \theta^2\, \text{Cl}_2(\theta )-\frac{128}{3}\, \theta\,  \text{Cl}_{-3}(\theta )+\frac{128}{3}\, \theta\,  \text{Cl}_3(\theta )\\
&\,+96\, \text{Cl}_{-4}(\theta )-72 \,\text{Cl}_4(\theta )-\frac{32  }{3}\,\zeta_3\, \theta-\frac{13 }{3}\,\theta^3\,.
\esp\eeq
The one-loop amplitude involving the operator $\cO_3$ is
\beq\bsp
\cA^{(0)}_3 &\,= m_t\,\Big[b_0 +4\log\tau+\eps\Big(b_1 + \log\tau\,b_0 + 2\,\log^2\tau\Big) \\
&\,+ \eps^2 \Big(b_2 + \log\tau\,b_1 +\frac{1}{2}\log^2\tau\,b_0 + \frac{2}{3}\log^3\tau\Big)+\ord(\eps^3)\Big]\,,
\esp\eeq
where the coefficients $b_i$ are given by
\begin{align}
b_0 &\,= \frac{2 \left(\tau -\theta ^2\right)}{\tau } -\frac{2\,\tau\, \theta}{\sqrt{(4-\tau ) \tau }}  \left(1-\frac{4}{\tau }\right)\,,\\
\nonumber b_1&\,= \frac{2}{\tau } \left[4\, \theta\,  \text{Cl}_{-2}(\theta )+8\, \text{Cl}_{-3}(\theta )+\theta ^2-6\, \zeta_3\right]+2 + \frac{\pi^2}{3} \\
\nonumber&\,+\frac{2\,\tau}{\sqrt{(4-\tau ) \tau }} \left(1-\frac{4}{\tau }\right) \,  \left[\theta\,  \log(4-\tau)-2\, \text{Cl}_{-2}(\theta )-\theta \right]\,,\\
\nonumber b_2&\,= \frac{1}{6 \tau }\big[48 \text{Cl}_{-2}(\theta )^2-96 \theta  \text{Cl}_{-2,-1}(\theta )-48 \theta  \text{Cl}_{-2}(\theta )-96 \text{Cl}_{-3}(\theta )-4 \theta ^4-\pi ^2 \theta ^2+12 \theta ^2\\
\nonumber&\,+ 72\, \zeta_3\big]+2+\frac{\pi ^2}{6}-\frac{4}{3}\,\zeta_3-\frac{\tau}{6 \sqrt{(4-\tau ) \tau }}\left(1-\frac{4}{\tau }\right)   \big[48 \,\text{Cl}_{-2,-1}(\theta )+6\, \theta\,  \log^2(4-\tau)\\
\nonumber&\,-12\, \theta\,  \log(4-\tau)+24\, \text{Cl}_{-2}(\theta )-24\, \text{Cl}_{-2}(\theta ) \,\log(4-\tau)+2 \theta ^3+\pi ^2\, \theta +12\, \theta \big]\,.
\end{align}
The finite remainder of the two-loop amplitude $\cA_3^{(1)}$ is
\begin{align}
\cR_3&\, = -i \pi\,  m_t\,\beta_0\, \left(1-\frac{\theta^2}{\tau }+2 \log\tau \right) + \frac{\theta ^2\, m_t}{\tau ^2}\,(16 \log\tau+35)\\
\nonumber&\,+\frac{m_t}{\tau } \Bigg[\theta \, \Big(-48\, \text{Cl}_{1,-2}(\theta )-\frac{8}{3} \,\text{Cl}_{2,1}(\theta )+\frac{1}{3} \,\text{Cl}_2(\theta )\, \log \tau -\frac{64}{3}\, \text{Cl}_{-2}(\theta )-74\, \text{Cl}_2(\theta )\Big)\\
\nonumber&\,\qquad-96\, \text{Cl}_{1,-3}(\theta )-48\, \text{Cl}_{2,-2}(\theta )-\frac{8}{3}\, \text{Cl}_{3,1}(\theta )-\frac{4}{3}\, \text{Cl}_3(\theta ) \,\log \tau +\frac{5}{3}\, \text{Cl}_2(\theta )^2\\
\nonumber&\,\qquad +\frac{61}{48}\, \theta ^4-\frac{2 \pi }{9}\,\theta ^3+\theta ^2 \left(\frac{1}{4}\,\log ^2\tau -\frac{92}{3}\,\log \tau +\frac{16}{3} \log (4-\tau )+5\,\zeta_2-\frac{100}{3}\right)\\
\nonumber&\,\qquad -\frac{64}{3}\, \text{Cl}_{-3}(\theta )-64\, \text{Cl}_3(\theta )-16\, \log\tau -\frac{104}{3}\, \zeta_3\,\log \tau +80\, \zeta_3-\frac{151}{3}\,\zeta_4-\frac{71}{3}\Bigg]\\
\nonumber&\,{ +m_t} \Bigg[\frac{32\,\theta}{3}\,  \left(2\, \text{Cl}_{-2}(\theta )- \text{Cl}_2(\theta )\right)+32\, \text{Cl}_{-3}(\theta )-16\, \text{Cl}_3(\theta )-8\, \zeta _3+\frac{5 }{3}\,\log ^2\tau +\frac{62}{3}\, \log \tau \\
\nonumber&\, \qquad-\theta^2\, \left(\frac{8}{3} \log (4-\tau )+\frac{4  }{3}\,\log\tau+\frac{1}{4}\right)+\frac{238}{3}\Bigg]\\
\nonumber&\,
-\frac{i \pi\,  \theta\,  (4-\tau ) \,m_t}{\sqrt{(4-\tau ) \tau }}\,\beta_0 + \frac{64\, \theta ^3\, m_t}{\tau ^2 \,\sqrt{(4-\tau ) \tau }} - \frac{2 \, m_t}{\sqrt{(4-\tau ) \tau }}\,\left(1-\frac{2}{\tau }\right)\,R(\theta) \\
\nonumber&\,+ \frac{\theta\,m_t}{6\,\sqrt{(4-\tau ) \tau }} \,\left[13\, \theta ^2+62-\frac{4}{\tau}  \,\left(63\, \theta ^2+62\right)\right] - \frac{(4-\tau) \,m_t}{\sqrt{(4-\tau ) \tau }}\Bigg[-\frac{32}{3} \text{Cl}_{-2}(\theta )+3 \text{Cl}_2(\theta )\\
\nonumber&\,\qquad+\theta\,  \left(\frac{16}{3}\, \log (4-\tau )-\frac{1}{6}\,\log\tau -\frac{71}{2}\right)\Bigg]\,.
\end{align}

Although the main focus of this paper is to include effects from dimension six operators that affect the gluon-fusion cross section through the top quark, let us conclude this section by making a comment about effects from the bottom, and to a lesser extent, the charm quark. The amplitudes presented in this section are only valid if the Higgs boson is lighter than the quark-pair threshold, $\tau<4$. It is, however, not difficult to analytically continue our results to the region above threshold where $\tau>4$. Above threshold, the variable $x$ defined in eq.~\eqref{eq:x_def_sqrt} is no longer a phase, but instead we have $-1<x<0$. As a consequence, the Clausen functions may develop an imaginary part. In the following we describe how one can extract the correct imaginary part of the amplitudes in the region above threshold (see also ref.~\cite{Graudenz:1992pv,Spira:1995rr,Anastasiou:2006hc,Anastasiou:2009kn}).

We start from eq.~\eqref{eq:clausen} and express all Clausen functions in terms of HPLs in $x$ and its inverse, e.g.,
\beq
\textrm{Cl}_{2}(\theta) =\textrm{Im}\,H(0,1;x) = \frac{1}{2i}\,\left[H(0,1;x) - H(0,1;1/x)\right]\,.
\eeq
HPLs evaluated at $1/x$ can always be expressed in terms of HPLs in $x$. For example, one finds
\beq
H(0,1;1/x) = -H(0,1;1/x) + i\pi\,H(0;x) - H(0,0;x)  +\frac{\pi^2}{3}\,,\qquad |x|=1 \textrm{~~and~~} \textrm{Im}\,x > 0\,.
\eeq
Similar relations can be derived for all other HPLs in an algorithmic way~\cite{Remiddi:1999ew,Maitre:2005uu,Maitre:2007kp}. The previous equation, however, is not yet valid above threshold, because the logarithms $H(0;x) = \log x$ may develop an imaginary part. Indeed, when crossing the threshold $x$ approaches the negative real axis from above, $x\to x+i 0$, and so the correct analytic continuation of the logarithms is
\beq\label{eq:cont_log}
H(0;x) = \log x \to H(0;-x) + i\pi\,.
\eeq
The previous rule is sufficient to perform the analytic continuation of all HPLs appearing in our results. Indeed, it is known that an HPL of the form $H(a_1,\dots,a_k;x)$ has a branch point at $x=0$ only if $a_k=0$, and, using the shuffle algebra properties of HPLs~\cite{Remiddi:1999ew}, any HPL of the form $H(a_1,\dots,a_k,0;x)$ can be expressed as a linear combination of products of HPLs such that if their last entry is zero, then all of its entries are zero. The amplitudes can therefore be expressed in terms of two categories of HPLs: those whose last entry is non-zero and so do not have a branch point at $x=0$, and those of the form $H(0,\dots,0;x)= \dfrac{1}{n!}\log^n x$, which are continued according to eq.~\eqref{eq:cont_log}.

Using the procedure outlined above, it is possible to easily perform the analytic continuation of our amplitudes above threshold. The resulting amplitudes contribute to the gluon-fusion process when light quarks, e.g., massive bottom and/or charm quarks, are taken into account. Hence, although we focus primarily on the effects from the top quark in this paper, our results can be easily extended to include effects from bottom and charm quarks as well.

\subsection{Renormalisation group running of the effective couplings}
After renormalisation, our amplitudes depend explicitly on the scale $\mu$, which in the following we identify with the factorisation scale $\mu_F$. It can, however, be desirable to choose different scales for the strong coupling constant, the top mass and the effective couplings. In this section we derive and solve the renormalisation group equations (RGEs) for these parameters.

\begin{figure}[h!]
  \begin{minipage}[t]{0.5\linewidth}
 \centering
 \includegraphics[width=.85\linewidth]{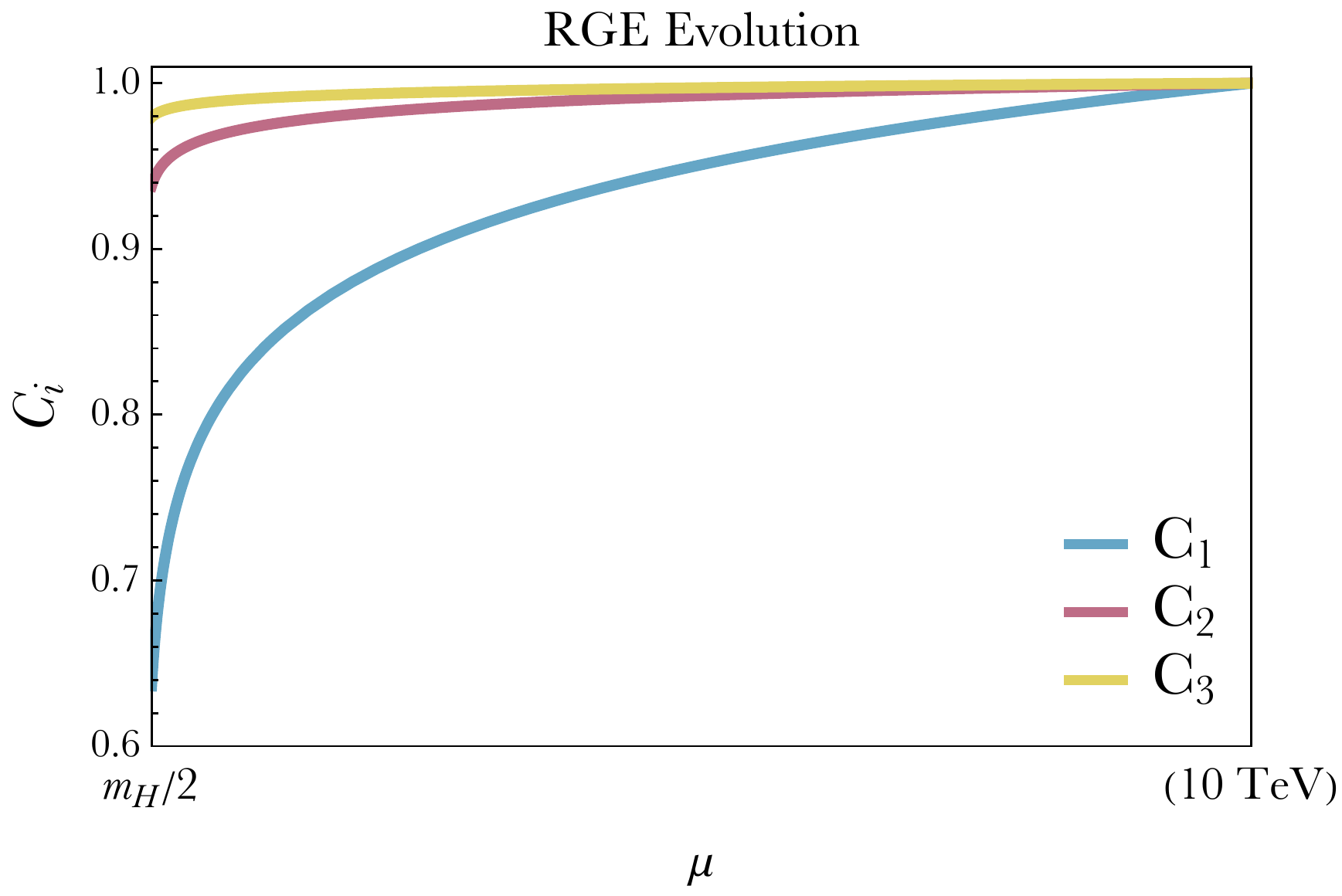}
 \end{minipage}
 \hspace{0.5cm}
  \begin{minipage}[t]{0.5\linewidth}
  \centering
  \includegraphics[width=.85\linewidth]{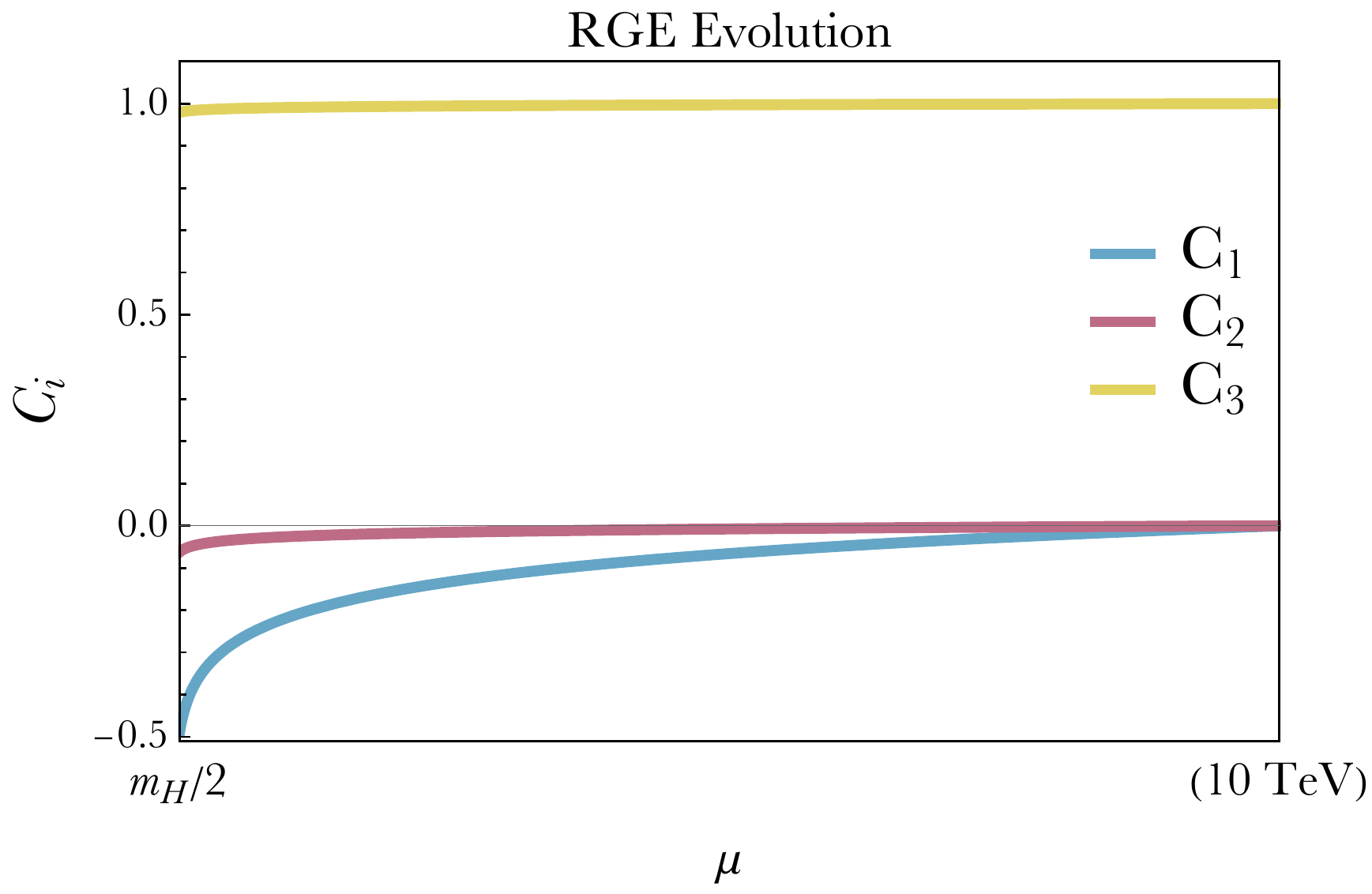}
  \end{minipage}
  \caption{\label{fig:running} Renormalization group evolution of the three Wilson coefficients between 10 TeV and $m_H/2$ in two scenarios. Left: $C_1=C_2=C_3=1$ at $\mu=10 \text{\ TeV}$. Right: $C_1=C_2=0$ and $C_3=1$ at $\mu=10 \text{\ TeV}$.}
\end{figure}

Since we are working in a decoupling scheme for the top mass, the RGEs for the strong coupling constant and the top mass are identical to the SM with $N_f=5$ massless flavours. We have checked that we correctly reproduce the evolution of $\alpha_s$ and $m_t$ in the $\overline{\textrm{MS}}$ scheme, and we do not discuss them here any further. For the RGEs satisfied by the effective couplings, we find
\beq\label{eq:C_RGE}
\frac{dC}{d\log\mu^2} = \gamma\,C\,,
\eeq
where $C=(C_1,C_2,C_3)^T$, and the anomalous dimension matrix is given by
\beq
\def\arraystretch{1.7}
\gamma= \begin{pmatrix}
  0 & 0 & 0 \\
  0 & 0 & \dfrac{1}{8 \pi^2\sqrt{2} }\dfrac{m_t(\mu^2)}{v}\\
  0 & 0 & 0
\end{pmatrix} + \dfrac{\alpha_s(\mu^2)}{\pi} \begin{pmatrix}
  -1 & 0 &  8 \dfrac{m_t(\mu^2)^2}{v^2} \\
  0 & 0 & \dfrac{23}{32\pi^2\sqrt{2}}\dfrac{m_t(\mu^2)}{v}\\
  0 & 0 & \dfrac{1}{6}
\end{pmatrix} + \ord(\alpha_s(\mu^2)^2)\,.
\eeq
As already mentioned in the previous section, the double pole from the two-loop counterterm in eq.~\eqref{eq:z23} cancels. We can solve the RGEs in eq.~\eqref{eq:C_RGE} to one loop, and we find
\beq\bsp
C_1(\mu^2)&\, = C_1(Q^2) -\frac{\alpha_s(Q^2)}{\pi}\,\log\frac{\mu^2}{Q^2}\,\left(C_1(Q^2) - 8\,C_3(Q^2)\,\frac{m_t^2(Q^2)}{v^2}\right) + \ord(\alpha_s(Q^2)^2)\,,\\
C_2(\mu^2)&\, = C_2(Q^2) + \sqrt{2}\,\frac{C_3(Q^2)}{16\,\pi^2}\,\log\frac{\mu^2}{Q^2}\,\frac{m_t(Q^2)}{v}\\
&\, -\sqrt{2}\,\frac{\alpha_s(Q^2)}{192\,\pi^3}\,C_3(Q^2)\,\log\frac{\mu^2}{Q^2}\,\frac{m_t(Q^2)}{v}\,\left(5\,\log\frac{\mu^2}{Q^2}-69\right) + \ord(\alpha_s(Q^2)^2)\,,\\
C_3(\mu^2)&\, = C_3(Q^2) +C_3(Q^2) \,\frac{\alpha_s(Q^2)}{6\pi} \,\log\frac{\mu^2}{Q^2}+ \ord(\alpha_s(Q^2)^2)\,.
\esp\eeq
We show in fig.~\ref{fig:running} the quantitative impact of running and mixing  by varying the renormalisation scale from 10 TeV to $m_H/2$ in two scenarios: one where all Wilson coefficients are equal at 10 TeV and another where only $C_3$ is non-zero. This latter example serves as a reminder of the need to always consider the effect of all the relevant operators in phenomenological analyses as choosing a single operator to be non-zero is a scale-dependent choice.

\section{Phenomenology}
\label{sec:pheno}
\subsection{Cross-section results}
In this section we perform a phenomenological study of Higgs production in the SMEFT, focusing on anomalous contributions coming from the top quark. Results are obtained within the {\sc MadGraph5\_aMC@NLO}  framework \cite{Alwall:2014hca}. The computation builds on the implementation of the dimension-six operators presented in ref.~\cite{Maltoni:2016yxb}. Starting from the SMEFT Lagrangian, all tree-level and one-loop amplitudes can be obtained automatically using a series of packages \cite{Alloul:2013bka, Degrande:2011ua,Degrande:2014vpa,
Hirschi:2011pa,Frederix:2009yq,Hirschi:2015iia}.  The two-loop amplitudes for the virtual corrections are implemented in the code through a reweighting method \cite{Maltoni:2014eza,Mattelaer:2016gcx}. Within the {\sc MadGraph5\_aMC@NLO} framework NLO results can be matched to
parton shower programs, such as  {\sc PYTHIA8} \cite{Sjostrand:2014zea} and
{\sc HERWIG++} \cite{Bahr:2008pv},  through the {\sc MC@NLO}
\cite{Frixione:2002ik} formalism.

Results are obtained for the LHC at 13 TeV with MMHT2014 LO/NLO PDFs \cite{Harland-Lang:2014zoa}, for
LO and NLO results respectively. The values of the input parameters are
\begin{flalign}
	&m_t=173\ \mathrm{GeV}\,, \quad
	m_H=125\ \mathrm{GeV}\,, \quad
	m_Z=91.1876\ \mathrm{GeV}\,, \\
	&\alpha_{EW}^{-1}=127.9\,, \quad
	G_F=1.16637\times10^{-5}\ \mathrm{GeV}^{-2}\,.
	\label{eq:input}
\end{flalign}
The values for the central scales for $\mu_R,\mu_F$ and $\mu_{EFT}$ are chosen as $m_H/2$, and we work with the top mass in the on-shell scheme.

We parametrise the contribution to the cross section from dimension-six operators as
\begin{flalign}
	\sigma=\sigma_{SM}+\sum_i\frac{1{\rm TeV}^2}{\Lambda^2}C_i\sigma_i
	+\sum_{i\leq j}
	\frac{1{\rm TeV}^4}{\Lambda^4}C_iC_j\sigma_{ij}.
	\label{eq:xsecpara}
\end{flalign}
  Within our setup we can obtain results for $\sigma_{SM}$, $\sigma_i$, and $\sigma_{ij}$. We note here that results for single Higgs and $H+j$ production in the SMEFT were presented at LO in QCD in ref.~\cite{Maltoni:2016yxb}. The normalisation of the operators used here differs from the one in ref.~\cite{Maltoni:2016yxb}, but we have found full agreement between the LO results presented here and those of ref.~\cite{Maltoni:2016yxb} when this difference is taken into account. Furthermore, the SM top-quark results obtained here have been cross-checked with the NLO+PS implementation of {\sc aMCSusHi} \cite{Mantler:2015vba}.

\begin{figure}[!]
	\centering
	\includegraphics[width=\textwidth]{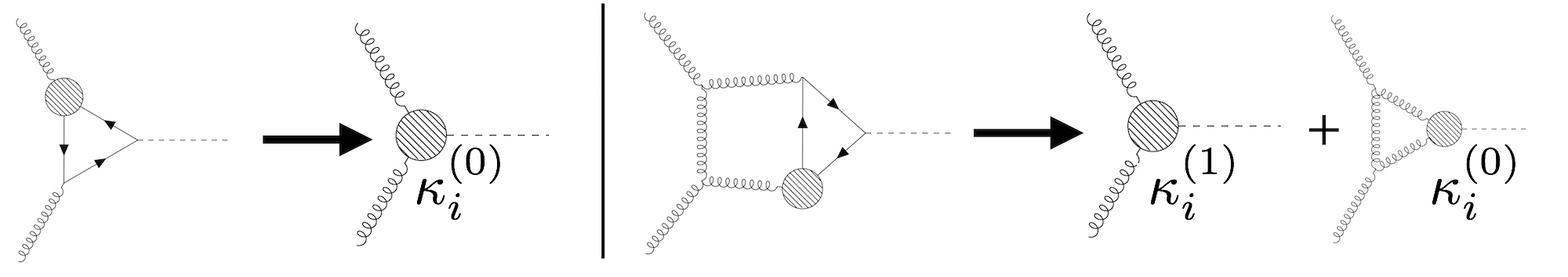}
	\caption{Diagrammatic description of the matching between the SMEFT and the infinite top mass EFT at LO (left) and at NLO (right). The NLO amplitude in the infinite top-mass EFT contains two elements: diagrammatic corrections, which contribute universally to the $K$-factors and Wilson coefficient corrections, which are non-universal.}
	\label{fig:nlomatching}
\end{figure}

Our results for the total cross section at the LHC at 13 TeV at LO and NLO are shown in table \ref{tab:tth1}. We include effects from bottom-quark loops (top-bottom interference and pure bottom contributions) into the SM prediction by using {\sc aMCSusHi}. However, in this first study, we neglect bottom-quark effects from dimension-six operators in $\sigma_i$ and $\sigma_{ij}$ as we assume them to be subleading. As mentioned above, our analytic results and MC implementation can be extended to also include these effects.
We see that the contributions from effective operators have $K$-factors that are slightly smaller then their SM counterpart, with a residual scale dependence that is almost identical to the SM. In the following we present an argument which explains this observation. We can describe the total cross section for Higgs boson production to a good accuracy by taking the limit of an infinitely heavy top quark, because most of the production happens near threshold. In this effective theory where the top quark is integrated out, all contributions from SMEFT operators can be described by the same contact interaction $\kappa G_{\mu\nu}^aG_a^{\mu\nu}H$. The Wilson coefficient $\kappa$ can be written as
\beq
\kappa=\kappa_0 + \sum C_i \kappa_i\,,
\eeq
where $\kappa_0$ denotes the SM contribution and $\kappa_i$ those corresponding to each operator ${\cal O}_i$ in the SMEFT. As a result each $\sigma_i$ is generated by the same Feynman diagrams both at LO and NLO in the infinite top-mass EFT.
The effect of radiative corrections is, however, not entirely universal as NLO corrections to the infinite top-mass EFT amplitudes come both from diagrammatic corrections and corrections to the Wilson coefficients $\kappa_i$, which can be obtained by matching the SMEFT amplitude to the infinite top mass amplitude, as illustrated in fig.~\ref{fig:nlomatching}.
Indeed, each $\kappa_i$ can be expressed in terms of SMEFT parameters as a perturbative series $\kappa_i = \kappa_i^{(0)} + \alpha_s \kappa_i^{(1)}+{\cal O}(\alpha_s^2)$. In the infinite top mass EFT, each $K$-factor $K_i$ can be decomposed as
\begin{equation}
K_i = K_U+ \alpha_s \frac{\kappa_i^{(1)}}{\kappa_i^{(0)}},
\end{equation}
where $K_U$ is the universal part of the $K$-factor, which is exactly equal to $K_2$. By subtracting $K_2$ to each $K_i$ in the infinite top mass limit numerically (setting $m_t=10{\rm TeV}$), we could extract the ratios $\alpha_s \frac{\kappa_i^{(1)}}{\kappa_i}$ and check explicitly that these non-universal corrections are subdominant compared to the universal diagrammatic corrections, which explains the similarity of the effects of radiative corrections for each contribution.

\begin{table}
\renewcommand{\arraystretch}{1.8}
 \makebox[\linewidth]{
	\begin{tabular}{llllll}
	\hline\hline
		13 TeV &$\sigma$ LO&$\sigma/\sigma_{SM}$ LO&
		$\sigma$ NLO&$\sigma/\sigma_{SM}$ NLO& $K$
		\\\hline
$\sigma_{SM}$& $21.3^{ +34.0+1.5\%}_{-25.0-1.5\%}$ & 1.0 & $36.6^{  +26.4+1.9\%}_{-20.0-1.6\%}$ & 1.0 & 1.71\\
$\sigma_{1}$& $-2.93^{ +34.0+1.5\%}_{-25.0-1.5\%}$ & -0.138 &$-4.70^{  +24.8+1.9\%}_{-20.0-1.6\%}$ & -0.127 &1.61\\
$\sigma_{2}$& $2660^{  +34.0+1.5\%}_{-25.0-1.5\%}$ & 125 & $4130^{  +23.9+1.9\%}_{-19.6-1.6\%}$& 114 & 1.55\\
$\sigma_{3}$& $50.5 ^{+34.0+1.5\%}_{-25.0-1.5\%}$ & 2.38 & $83.5^{+26.0+1.9\%}_{-20.6-1.6\%}$  & 2.28 & 1.65 \\
$\sigma_{11}$& $0.0890 ^{+34.0+1.5\%}_{-25.0-1.5\%}$ & 0.0042 & $0.141^{+24.8+1.9\%}_{-20.0-1.6\%}$ & 0.0038 & 1.59\\
$\sigma_{22}$&$74100^{  +34.0+1.5\%}_{-25.0-1.5\%}$ & 3480 &$109100^{  +22.6+1.9\%}_{-18.9-1.6\%}$ & 3000 &1.47\\
$\sigma_{33}$& $26.6 ^{+34.0+1.5\%}_{-25.0-1.5\%}$ & 1.25 &$41.6^{+25.3+2.0\%}_{-20.4-1.7\%}$ & 1.13 &1.56 \\
$\sigma_{12}$& $ -162^{+34.0+1.5\%}_{-25.0-1.5\%}$ & -7.61 & $-248^{+23.6+1.9\%}_{-19.5-1.6\%}$ & -6.78 & 1.53\\
$\sigma_{13}$&$-3.08^{  +34.0+1.5\%}_{-25.0-1.5\%}$ & -0.145 &$-5.04^{  +25.4+1.9\%}_{-20.3-1.6\%}$ &  -0.138 &1.64\\
$\sigma_{23}$& $ 2800^{+34.0+1.5\%}_{-25.0-1.5\%}$ & 131  &$4460^{+24.6+1.9\%}_{-19.9-1.6\%}$ &122 & 1.59\\
\hline
	\end{tabular}}
\caption{\label{tab:tth1}
	Total cross section in pb for $pp\to H$ at 13 TeV, as
	parametrised in eq.~(\ref{eq:xsecpara}).}
\end{table}
Our results can be used to put bounds on the Wilson coefficients from measurements of the gluon-fusion signal strength $\mu_{ggF}$ at the LHC. Whilst here we do not attempt to perform a rigorous fit of the Wilson coefficients, useful information can be extracted by a simple fit. For illustration purposes, we use the recent measurement of the gluon-fusion signal strength in the diphoton channel by the CMS experiment~\cite{CMS:2017rli}
\begin{equation}
	\mu_{ggF} = 1.1 \pm 0.19,
\end{equation}
which we compare to our predictions for this signal strength under the assumption that the experimental selection efficiency is not changed by BSM effects
\begin{equation}
	\mu_{ggF} = 1 + \left(\frac{C_1 \sigma_1 + C_2 \sigma_2 + C_3 \sigma_3}{\sigma_{SM}}\right),
\end{equation}
where we set $\Lambda=1 {\rm TeV}$ and kept only the $\ord(1/\Lambda^2)$ terms. We therefore find that we can put the following constraint on the Wilson coefficients with 95\% confidence level:
\begin{equation}
-0.28< -0.128 C_1 + 114 C_2 + 2.28 C_3 < 0.48.
\label{linear}
\end{equation}
While the correct method for putting bounds on the parameter space of the SMEFT is to consider the combined contribution of all relevant operators to a given observable, the presence of unconstrained linear combinations makes it interesting to consider how each operator would be bounded if the others were absent in order to obtain an estimate of the size of each individual Wilson coefficient. Of course such estimates must not be taken as actual bounds on the Wilson coefficients and should only be considered of illustrative value. We obtain
\begin{equation}
-3.8<C_1<2.2,\, -0.0025 < C_2 < 0.0043,\, -0.12 < C_3 < 0.21\,.
\end{equation}
For these individual operator constraints, the impact of the $\sigma_{ii}$ terms on the limits is at most 10\%.

For reference we note that if one includes the $\ord(1/\Lambda^4)$ contributions the linear combination in the bound becomes a quadratic one:
\begin{eqnarray}\nonumber
-0.28&<& -0.128 C_1 + 114 C_2 + 2.28 C_3 + 0.0038 C_1^2+ 3000 C_2^2\\
 &&+1.13 C_3^2 - 6.78 C_1 C_2-0.138 C_1 C_3  +122 C_2 C_3 < 0.48\,. \label{limit}
\end{eqnarray}

\subsection{Differential distributions}
In the light of differential Higgs measurements at the LHC, it is important to examine the impact of the dimension-six operators on the Higgs $p_T$ spectrum. It is known that measurements of the Higgs $p_T$ spectrum can be used to lift the degeneracy between $\cO_{1}$ and $\cO_{2}$ \cite{Grojean:2013nya,Buschmann:2014sia,Maltoni:2016yxb}. For a realistic description of the $p_T$ spectrum, we match our NLO predictions to the parton shower with the MC@NLO method \cite{Frixione:2002ik}, and we use PYTHIA8 \cite{Sjostrand:2014zea} for the parton shower. Note that we have kept the shower scale at its default value in MC@NLO, which gives results that are in good agreement with the optimised scale choice of ref.~\cite{Harlander:2014uea}, as discussed in ref.~\cite{Mantler:2015vba}.

\begin{figure}[h!]
 \begin{minipage}[t]{0.46\linewidth}
\centering
\includegraphics[width=.99\linewidth,trim= 2cm 8cm 2cm 2cm]{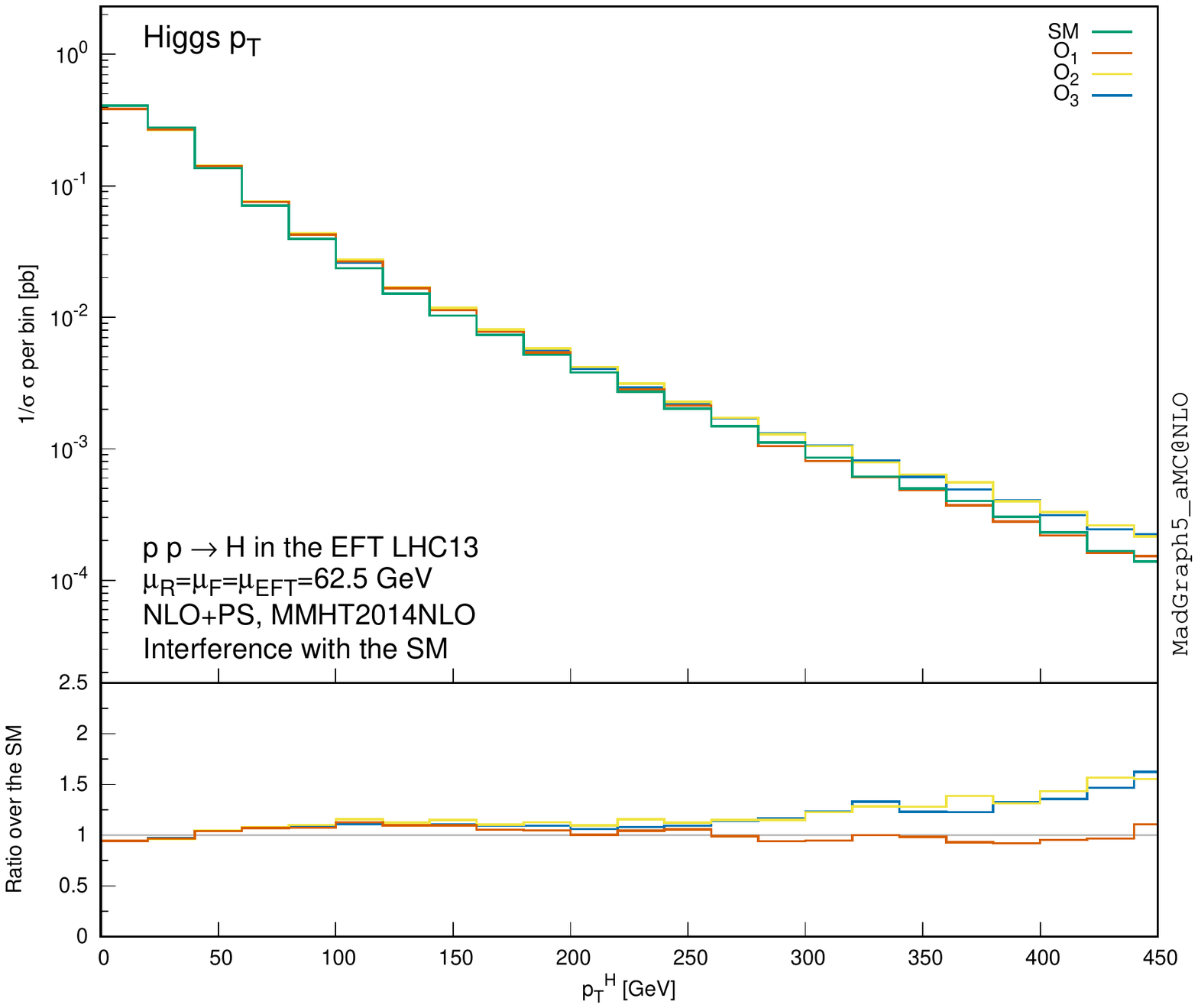}
\end{minipage}
\hspace{0.5cm}
 \begin{minipage}[t]{0.46\linewidth}
\centering
\includegraphics[width=.99\linewidth,trim= 2cm 8cm 2cm 2cm]{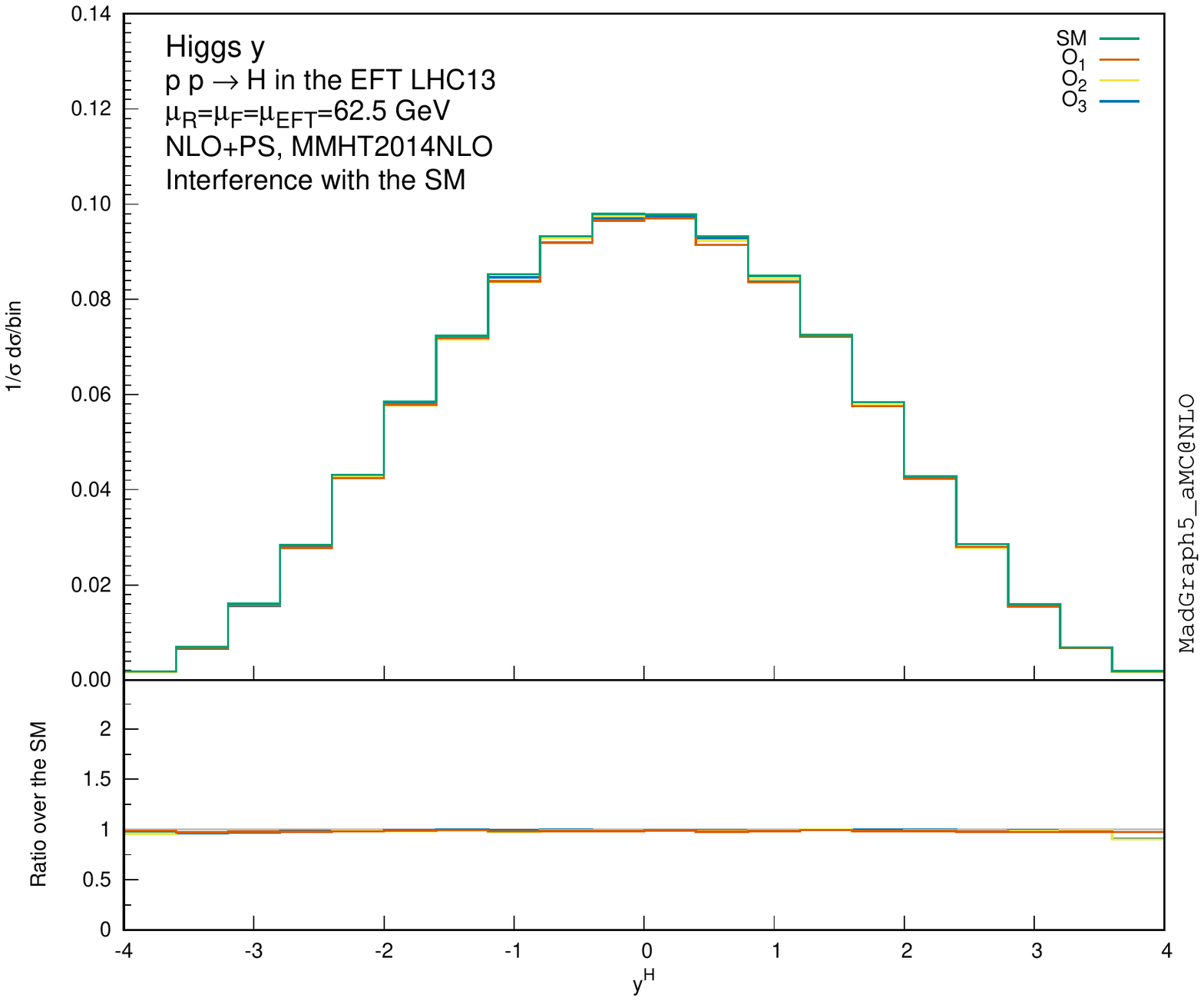}
\end{minipage}
\caption{\label{fig:ptH}
Higgs distributions, normalised for the interference contributions from $\sigma_i$.  Left:
Higgs transverse momentum. Right: Higgs rapidity.  SM contributions and individual operator contributions are
displayed.  Lower panels give the ratio over the SM.}
\end{figure}

\begin{figure}[h!]
 \begin{minipage}[t]{0.46\linewidth}
 \centering
 \includegraphics[width=.99\linewidth, trim= 2cm 8cm 2cm 4cm]{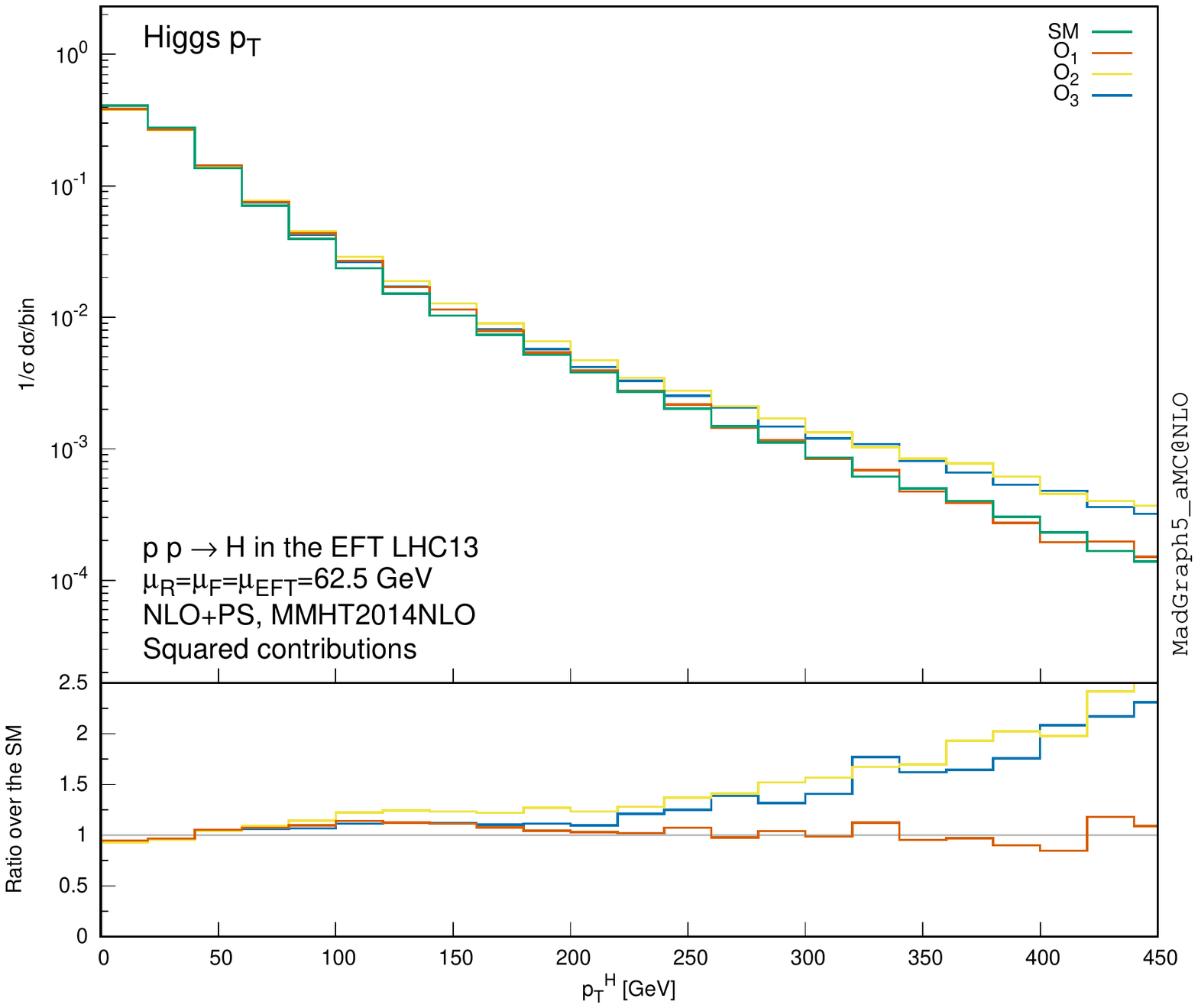}
 \end{minipage}
\hspace{0.5cm}
 \begin{minipage}[t]{0.46\linewidth}
 \centering
 \includegraphics[width=.99\linewidth, trim= 2cm 8cm 2cm 4cm]{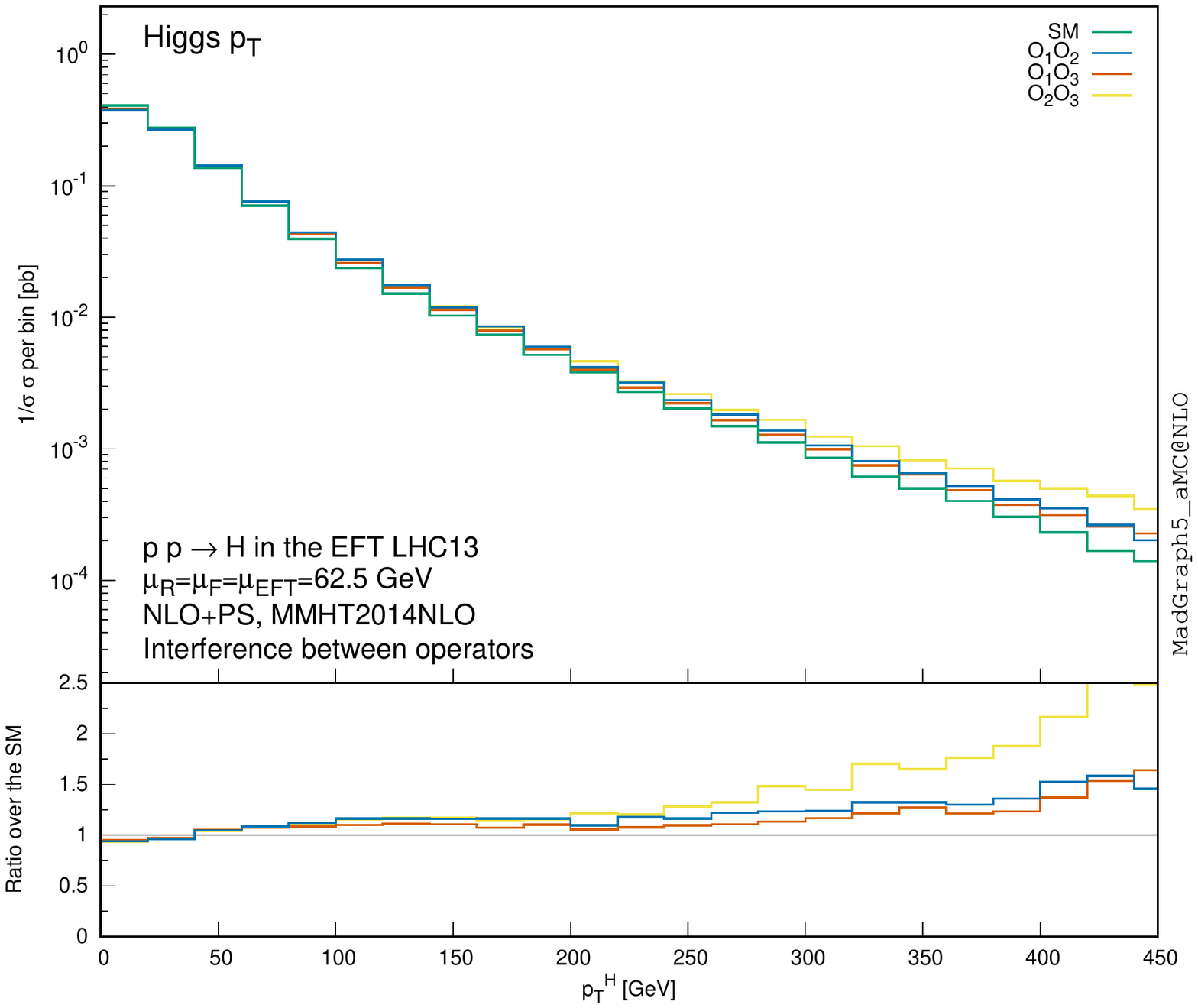}
 \end{minipage}
\caption{\label{fig:squared}
Higgs transverse momentum distributions, normalised.  Left: squared contributions
$\sigma_{ii}$. Right: interference between operators, $\sigma_{ij}$.  SM contributions and operator contributions are
displayed.  Lower panels give the ratio over the SM.}
\end{figure}

The normalised distributions for the transverse momentum and rapidity of the Higgs boson are shown in figs.~\ref{fig:ptH} for the interference contributions. The impact of the $\ord(1/\Lambda^4)$ terms is demonstrated in fig.~\ref{fig:squared} for the transverse momentum distribution. We find that the operators $\cO_{3}$ and $\cO_{2}$ give rise to harder transverse momentum tails, while for $\cO_{1}$ the shape is identical to the SM. The dimension-six operators have no impact on the shape of the rapidity distribution. The $\ord(1/\Lambda^4)$ contributions involving $\cO_{3}$ and $\cO_{2}$ are harder than those involving $\cO_{1}$.

\begin{figure}[h!]
\centering
\includegraphics[width=.7\linewidth,trim= 2cm 8cm 0cm 2cm]{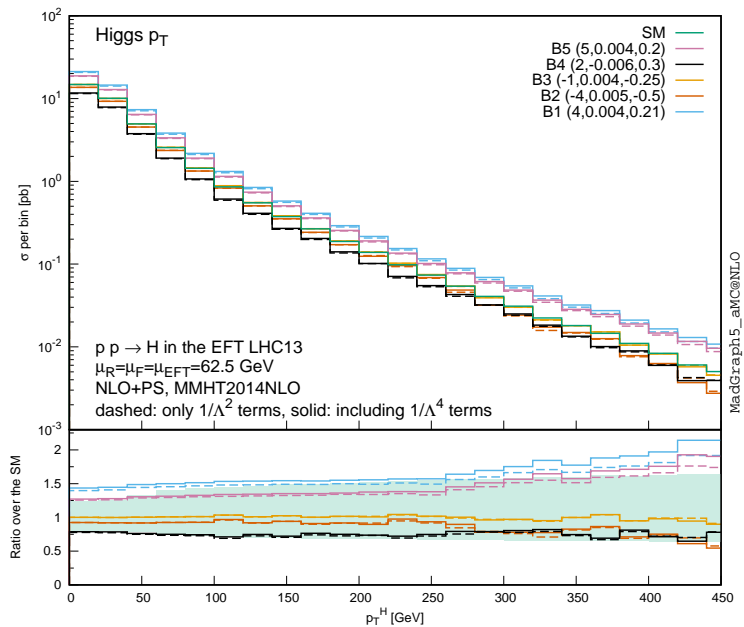}
\caption{\label{fig:bench}
Transverse momentum distributions of the Higgs for different values of the Wilson coefficients. The lower panel shows the ratio over the SM prediction for the various benchmarks and the SM scale variation band. }
\end{figure}

Finally we show the transverse momentum distributions for several benchmark points which respect the total cross-section bounds in fig.~\ref{fig:bench}. The operator coefficients are chosen such that eq.~\eqref{limit} is satisfied. We find that larger deviations can be seen in the tails of the distributions for coefficient values which respect the total cross-section bounds.

\subsection{Renormalisation group effects}

\begin{figure}[h!]
\centering
\includegraphics[width=1.3\linewidth,trim= 0cm 14cm 2cm 4cm]{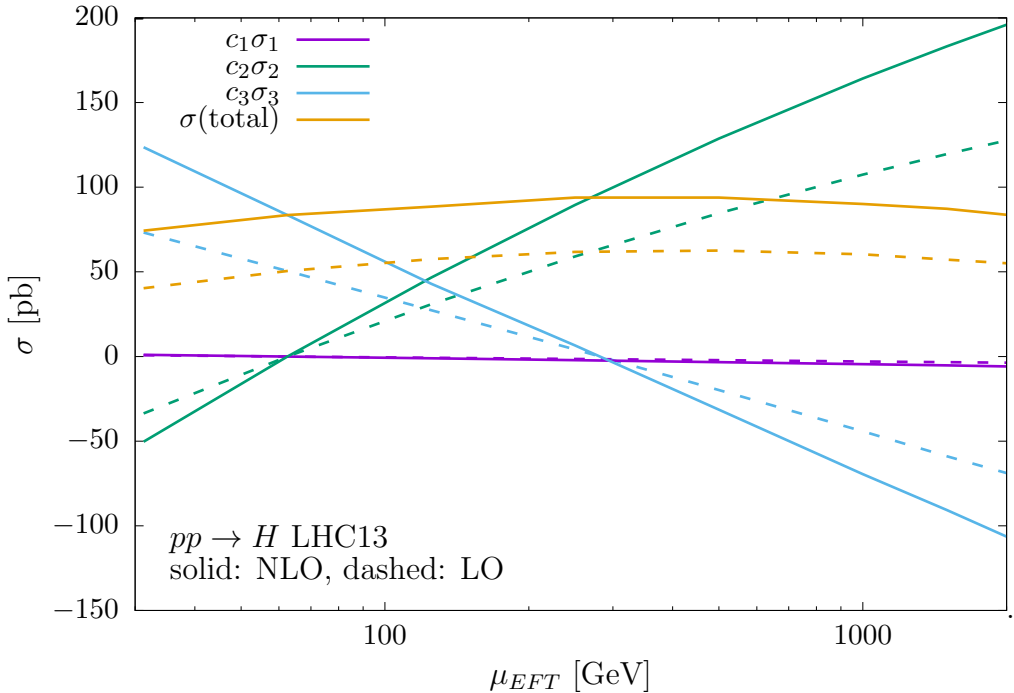}
\caption{\label{fig:RG}
Contributions of the three operators to the inclusive Higgs production cross section at the LHC at 13 TeV as a function of the EFT scale. Starting from one non-zero coefficient at $\mu_{EFT}=m_H/2$ we compute the EFT contributions at different scales, taking into account the running and mixing of the operators. LO and NLO predictions are shown in dashed and solid lines respectively.}
\end{figure}

The impact of running and mixing between the operators is demonstrated in fig.~\ref{fig:RG}, where we show the individual ($\ord(1/\Lambda^2)$) contributions from the three operators in gluon-fusion Higgs production at LO and NLO, as a function of $\mu_{EFT}$, assuming that $C_{3}=1$, $C_{1}=C_{2}=0$ at $\mu_{EFT}=m_H/2$ and $\Lambda=1$ TeV. While at $\mu=m_H/2$ the only contribution is coming from the chromomagnetic operator, this contribution changes rapidly with the scale. While the effect of the running of $C_{3}$ is only at the percent level, $\sigma_{3}$ has a strong dependence on the scale. At the same time non-zero values of $C_{1}$ and $C_{2}$ are induced through renormalisation group running, which gives rise to large contributions from $\cO_{2}$. We find that the dependence on the EFT scale is tamed when the sum of the three contributions is considered. This is the physical cross section coming from $C_{3}(m_H/2)=1$ which has a weaker dependence on the EFT scale. The dependence of this quantity on the scale gives an estimation of the higher order corrections to the effective operators and should be reported as an additional uncertainty of the predictions. By comparing the total contributions at LO and NLO we find that the relative uncertainty is reduced at NLO.

\section{Conclusion and outlook}
\label{sec:conclusion}

A precise determination of the properties of the Higgs boson and, in particular, of its couplings to the other SM particles is one of the main goals of the LHC programme of the coming years. The interpretation of such measurements, and of possible deviations in the context of an EFT, allows one to put constraints on the type and strength of hypothetical new interactions, and therefore on the scale of new physics, in a model-independent way.  The success of this endeavour will critically depend on having theoretical predictions that at least match the precision of the experimental measurements, both in the SM and in the SMEFT. 

In this work we have computed for the first time the contribution of the ($CP$-even part of the) $ \bar Q_L \Phi  \sigma q_R G$ operator to the inclusive Higgs production at NLO in QCD. Since the NLO corrections for the other two ($CP$-even) operators entering the same process are available in the literature, this calculation completes the SMEFT predictions for this process at the NLO accuracy. Even though our results can be easily extended to include anomalous couplings of the bottom quark, we have considered in the detail the case where new physics mostly affects  the top-quark couplings. Our results confirm the expectations based on previous calculations and on the general features of gluon-fusion Higgs production: at the inclusive level the $K$-factor is of the same order as that of the SM and of the other two operators. The residual uncertainties estimated by renormalisation and factorisation scale dependence also match extremely well. The result of the NLO calculation confirms that the chromomagnetic operator cannot be neglected for at least two reasons.  The first is of purely theoretical nature: the individual effects of $ \bar Q_L \tilde \Phi \sigma t_R G$ and $\Phi^\dagger \Phi GG$ are very much dependent on the EFT scale, while their sum is stable and only mildly affected by the scale choice.  The second draws from the present status of the constraints. Considering the uncertainties in inclusive Higgs production cross section measurements and the constraints from $t\bar t$ production, the impact of the chromomagnetic operator cannot be neglected in global fits of the Higgs couplings. As a result, a two-fold degeneracy is left unresolved by a three-operator fit using the total Higgs cross section and one is forced to look for other observables or processes to constrain all three of the operators. 

The implementation of the finite part of the two-loop virtual corrections into {\sc MadGraph5\_aMC@NLO} has also allowed us to study the process at a fully differential level, including the effects of the parton shower resummation and in particular to compare the transverse momentum distributions of the SM and the three operators in the region of the parameter space where the total cross section bound is respected. Once again, we have found that the contributions from $ \bar Q_L \tilde \Phi \sigma t_R G$ and $\Phi^\dagger \Phi GG$ are similar and produce a shape with a harder tail substantially different from that of the SM and the Yukawa operator (which are the same). While $ \bar Q_L  \tilde \Phi \sigma t_R G$ and $\Phi^\dagger \Phi GG$ cannot really be distinguished in gluon-fusion Higgs production, they do contribute in a very different way to $t\bar t H$ where the effect of $\Phi^\dagger \Phi GG$ is extremely weak. Therefore, we expect that $H, H$+jet, and $t \bar t H$ (and possibly $t\bar t$) can effectively constrain the set of the three operators. 

In this work we have mostly focused our attention on the top-quark-Higgs boson interactions and only considered $CP$-even operators. As mentioned above and explained in section~\ref{sec:virtuals}, extending it to include anomalous couplings for lighter quarks, the bottom and possibly the charm, is straightforward. On the other hand, extending it to include $CP$-odd operators requires a new independent calculation. We reckon both developments worth pursuing.

\section*{Acknowledgements}
This work was supported in part by the ERC grant ``MathAm'', by the FNRS-IISN convention "Fundamental Interactions" FNRS-IISN 4.4517.08, and by the European Union Marie Curie Innovative Training Network MCnetITN3 722104.   E.V.~is supported by the research programme of the Foundation for Fundamental Research on Matter (FOM), which is part of the Netherlands Organisation for Scientific Research (NWO). ND acknowledges the hospitality of the CERN TH department while this work was carried out.

\bibliographystyle{JHEP}
\bibliography{refs}

\end{document}